\documentclass[preprint,12pt]{elsarticle}

\usepackage[utf8]{inputenc}
\usepackage[T1]{fontenc}
\usepackage{lmodern}
\usepackage{amsmath,amssymb,amsthm}
\usepackage{booktabs,tabularx,array}
\usepackage{enumitem}
\usepackage{geometry}
\usepackage[hidelinks,hypertexnames=false]{hyperref}
\usepackage{natbib}
\usepackage{float}
\usepackage{graphicx}
\usepackage{tikz}
\usetikzlibrary{arrows.meta,positioning}

\theoremstyle{remark}

\journal{Computers \& Security}

\begin{document}

\begin{frontmatter}

\title{Mean Time to Remediate Is Not a Fielding Model:\\
A Cadence Audit for Enterprise Vulnerability Management}

\author{Alexander Omelchenko}
\ead{aomelchenko@constructor.university}
\address{Constructor University Bremen gGmbH,
Campus Ring 1, 28759 Bremen, Germany}

\begin{abstract}
Enterprise security teams commonly summarize remediation through mean time to
remediate (MTTR), SLA compliance, dwell time, or detection delay.  These metrics
are useful, but they can hide how fixes actually reach the estate: continuously,
through scheduled maintenance windows, in deployment rings, or through emergency
bypass paths.  This paper introduces a remediation-cadence audit for enterprise
vulnerability management.  The audit records routine mean lag, release period,
release fraction, cohort geometry, emergency/routine split, non-fielding delay,
local residual-pressure evidence, and declared rate scenario.  It compares a
continuous same-mean shortcut with the recorded release calendar and reports a
local capacity verdict plus a calendar discount: the fraction of mean-only local
capacity consumed by calendarized fielding.  Worked notional packets with the
same 30-day mean lag show why this matters.  Under the normalized screening
scenario, a two-month release train consumes 17.4\% of mean-only capacity, a
monthly train 5.2\%, and a two-week screen 1.3\%; across a 16-fold
attacker-adjustment rate band, the two-month discount remains at least about
12\% and the monthly discount stays in the resolution-sensitive 3--8\% range.
The audit therefore turns cadence assessment into an evidence-resolution
question: when the discount is material relative to residual-pressure
uncertainty or claimed headroom, MTTR/SLA should not be used alone as fielding
evidence.  Release-geometry checks show that deployment rings do not
automatically recover the continuous benchmark, and cohort staggering can help
or hurt near capacity.  The result is a reproducible governance diagnostic, not
a breach predictor or CVE prioritizer.
\end{abstract}


\end{frontmatter}

\section{Introduction}
\label{sec:introduction}

A vulnerability-management dashboard can improve while the underlying fielding
process becomes less safe to summarize.  An enterprise security team may report
lower mean time to remediate (MTTR), better SLA compliance, and shorter detection
delay, while at the same time changing how defensive updates reach the estate:
routine fixes are held for scheduled maintenance windows, assets move through
deployment rings, and high-severity issues bypass the ordinary process.  From a
dashboard perspective the program looks healthier.  From a security-operations
perspective, however, the relevant question is different: is the reported mean
lag still a safe representation of how defensive changes are actually fielded?

This paper studies that question as a remediation-cadence audit.  The audit
addresses a narrow but important reporting failure: MTTR, SLA compliance, MTTD,
and dwell-time measures are useful governance indicators, but they are not
fielding models.  Two remediation processes can have the same reported mean lag
while exposing the estate to different release geometry.  One process may field
changes continuously; another may wait for monthly, six-week, bimonthly, or
quarterly release windows; another may clear only part of the eligible backlog
at each window; and another may split emergency and routine fixes into different
channels.  The mean lag summarizes average timing.  It does not say whether the
release calendar itself changes the local security-governance conclusion.

The operational setting is familiar.  Enterprise patch-management guidance
treats remediation as a process of identifying, prioritizing, acquiring,
installing, and verifying updates \citep{nistPatchMgmt}.  Public vendor and
cloud documentation describes monthly security-update cycles, scheduled patching,
maintenance configurations, and out-of-band update paths
\citep{msReleaseCycle,azureUpdateManager}.  CISA's KEV catalog and current
risk-based remediation directives illustrate exploited-vulnerability governance
and remediation-deadline objects \citep{cisaKEV,cisaBOD2604}.  Remediation
measurement studies further show that vulnerability-remediation data can be
skewed, long-tailed, and poorly summarized by a single mean \citep{cyentiaP2P}.
These sources motivate the timing side of the audit.  They do not, by
themselves, say whether a mean remediation lag is a safe fielding representation
for a local adaptive-risk claim.

The audit therefore records two kinds of evidence.  The first is timing and
fielding evidence: routine mean lag, release-window period, per-window release
fraction, cohort phases or deployment rings, emergency-bypass rules, and a
non-fielding delay budget for detection, validation, approval, testing,
packaging, and other hard delays.  The second is local residual-pressure
evidence: an interval derived from BAS, red-team, purple-team, control-validation,
postmortem, or expert-scored evidence for the audited channel.  Public CVE,
CVSS, EPSS, KEV, OSV, or advisory data can help define scope and prioritization,
but they cannot replace local evidence about how the organization's controls,
detections, response processes, and compensating measures perform on the
channel being audited \citep{firstCVSS,jacobsEPSS,cisaKEV,mitreATTACK}.

The audit returns a local capacity verdict and a calendar discount.  A local
channel is inside capacity when small disturbances in intended defensive posture,
fielded defensive posture, and attacker technique share decay under the recorded
remediation process.  It is outside capacity when those disturbances grow from
release window to release window: attacker adjustment outpaces the fielding of
defensive changes on that channel.  The calendar discount reports how much of
the mean-only local capacity is consumed by the recorded release calendar under
a declared local-channel rate scenario.  This second output is crucial.  It
means the audit does not depend only on a sharp point estimate of residual
pressure.  If the discount is material relative to the claimed headroom, or if
the residual-pressure evidence is too coarse to resolve the discount, the
governance finding is that MTTR/SLA should not be used alone as fielding
evidence.

The worked packets in this paper use a 30-day routine mean lag and a 15-day
non-fielding delay budget.  Under the normalized screening scenario used for the
main calculations, a coarse two-month release train with full eligible backlog
clearing consumes about 17.4\% of mean-only local capacity.  A monthly train with
the same mean lag consumes about 5\%.  A shorter two-week screen consumes about
1.3\%.  Across the rate-scenario band checked in Section~\ref{sec:validation-robustness},
the two-month discount remains at least about 12\%, the monthly discount stays
in the 3--8\% resolution-sensitive band, and the two-week screen remains below
2\%.  These values change the interpretation of the worked example.  The
two-month packet gives a resolved cadence warning that remains visible under
coarser evidence intervals.  The monthly packet is still important because it is
operationally common, but it is more resolution-sensitive: a binary
inside/outside verdict requires sharper local residual-pressure evidence.  The
audit therefore reports not only whether a point estimate falls between two
capacity boundaries, but also how much evidence resolution is needed before a
mean-only fielding claim is defensible.

The release-geometry checks add two further operational findings.  Splitting a
monthly train into deployment rings at fixed per-asset cadence does not recover
the continuous same-mean benchmark; it approaches a phase-averaged calendar
process.  Cohort staggering is also not automatically stabilizing near capacity:
depending on the local channel and phase schedule, staggering can help or hurt.
Thus rings and cohort phases should be recorded as audit fields rather than used
as verbal assurance that staged rollout behaves like continuous fielding.

The technical backend extends an implementation-filter mechanism, but the
contribution here is the security-operations audit built on top of it
\citep{omelchenkoCNSNS}.  The paper contributes four elements.  First, it
formulates a remediation-cadence audit record that joins timing evidence,
release geometry, emergency/routine channel separation, hard-delay budget,
local residual-pressure evidence, and rate-scenario assumptions in one
governance object.  Second, it introduces calendar discount as a practical
measure of how much capacity is consumed by calendarized fielding.  Third, it
uses notional but reproducible worked packets to show how the same reported mean
lag can produce different evidence-resolution and capacity readings under
different release calendars.  Fourth, it validates release-geometry effects
showing that rings do not automatically recover the continuous benchmark and
that cohort phasing can change the local verdict near capacity.

The remainder of the paper follows the audit rather than the mathematical
machinery.  Section~\ref{sec:audit} defines the remediation-cadence workflow,
audit record, output statuses, and calendar-discount interpretation.
Section~\ref{sec:worked-audit} gives worked notional audit packets and shows how
calendar discount changes the management reading.  Section~\ref{sec:model}
defines the calculation interface from an audit packet to capacity boundaries
and discount.  Section~\ref{sec:calendarization-penalty} explains why release
windows create a calendar discount even when the mean lag is matched.
Section~\ref{sec:validation-robustness} checks the finite thresholds,
hard-delay placement checks, deployment rings, and cohort-phasing effects.  The final
sections position the audit relative to vulnerability-management and security
measurement work, state limitations and responsible use, and summarize the
implications for enterprise remediation reporting.

\section{Remediation-cadence audit workflow}
\label{sec:audit}

A remediation-cadence audit is a security-operations check for one question:
when MTTR, SLA compliance, or a mean remediation lag is used as evidence that a
local adaptive-risk channel is manageable, is the mean-only fielding
representation adequate, or does the actual release calendar consume a material
part of the available local capacity?  The audit is not a general
vulnerability-prioritization method.  Its purpose is narrower: to decide whether
the remediation calendar itself must be disclosed, tested, and governed as a
security-relevant fielding variable.

The audit uses the terms \emph{inside capacity} and \emph{outside capacity} in
an operational sense.  A local channel is inside capacity when small disturbances
in intended defensive posture, fielded defensive posture, and attacker technique
share decay under the recorded remediation process.  It is outside capacity when
those disturbances grow from release window to release window: attacker
adjustment outpaces the fielding of defensive changes on that channel.  The
verdict is therefore a local fielding-governance statement, not a forecast that a
specific incident will occur.

The audit deliberately returns two outputs.  The first is the familiar
inside/outside capacity verdict obtained by comparing a local residual-pressure
interval \(I_L=[L_-,L_+]\) with the mean-only and calendar-aware capacity
boundaries.  The second is the \emph{calendar discount}
\begin{equation}
    \Delta_{\mathrm{cal}}
    =
    \frac{
    L_{\mathrm{crit}}^{\mathrm{cont}}
    -
    L_{\mathrm{crit}}^{\mathrm{cal}}
    }{
    L_{\mathrm{crit}}^{\mathrm{cont}}
    } .
\label{eq:audit-calendar-discount}
\end{equation}
The discount is the fraction of mean-only local capacity consumed by the
recorded release calendar.  This number is reported even when the evidence for
\(L\) is too coarse for a sharp inside/outside verdict.  In practical terms, if a
team's claimed residual-pressure headroom is smaller than the calendar discount,
then MTTR/SLA alone is not a safe fielding representation for that local claim.
If the uncertainty in \(L\) is larger than the discount, the correct audit
status is not a forced warning or a forced pass, but an input-resolution finding.

The discount should be interpreted under a declared local-channel rate scenario.
The default calculation in this paper is a normalized screening convention in
which local defensive-revision and attacker-adjustment rates are measured in
units of the routine mean lag.  A production audit should either declare that it
is using this screening convention, supply local rate estimates, or run scenario
bands over plausible rates.  If the calendar discount is stable across the
scenario band, the cadence finding is robust.  If it is not stable, the audit
should report a rate-scenario limitation rather than a sharp cadence verdict.

The audit should be run when the remediation process changes, or when the
security claim being made depends on a mean lag rather than on fielding
evidence.  Typical triggers include migration from rolling deployment to monthly,
six-week, bimonthly, or quarterly release trains; new change-freeze periods or
approval gates; introduction of deployment rings or phase-shifted cohorts;
emergency bypass being carved out of the routine path; SLA improvement
accompanied by persistent incident, BAS, red-team, or control-validation
pressure; BAS evidence that attacker activity is substituting toward techniques
not yet covered by routine remediation; major endpoint, cloud, or application
deployment tooling changes; and repeated exceptions to the nominal release
calendar.  The common feature is that MTTR may remain stable, or even improve,
while the timing structure of defensive fielding has changed.

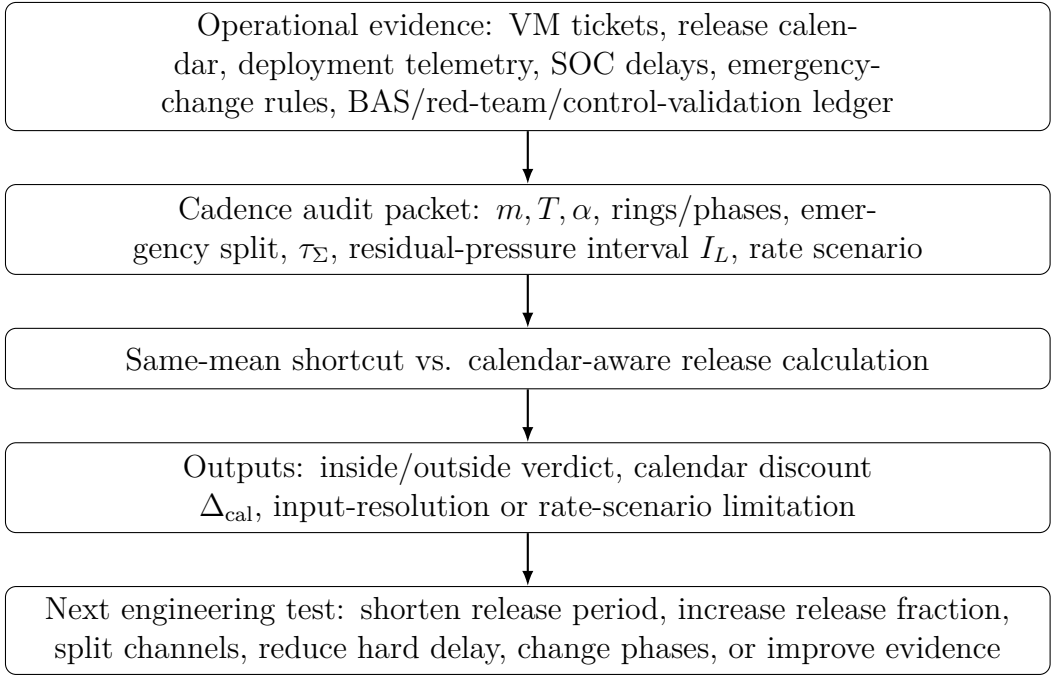
\begin{figure}[H]
\centering
\begin{tikzpicture}[
    node distance=7mm,
    box/.style={
        draw,
        rounded corners,
        align=center,
        text width=0.82\textwidth,
        minimum height=8mm,
        inner sep=4pt
    },
    arrow/.style={-{Latex[length=2mm]}, thick}
]
\node[box] (evidence)
{Operational evidence: VM tickets, release calendar, deployment telemetry,
SOC delays, emergency-change rules, BAS/red-team/control-validation ledger};

\node[box, below=of evidence] (packet)
{Cadence audit packet: \(m,T,\alpha\), rings/phases, emergency split,
\(\tau_\Sigma\), residual-pressure interval \(I_L\), rate scenario};

\node[box, below=of packet] (calc)
{Same-mean shortcut vs. calendar-aware release calculation};

\node[box, below=of calc] (outputs)
{Outputs: inside/outside verdict, calendar discount
\(\Delta_{\mathrm{cal}}\), input-resolution or rate-scenario limitation};

\node[box, below=of outputs] (action)
{Next engineering test: shorten release period, increase release fraction,
split channels, reduce hard delay, change phases, or improve evidence};

\draw[arrow] (evidence) -- (packet);
\draw[arrow] (packet) -- (calc);
\draw[arrow] (calc) -- (outputs);
\draw[arrow] (outputs) -- (action);
\end{tikzpicture}
\caption{Remediation-cadence audit workflow.  The audit does not treat MTTR as
a fielding model.  It records the release process, compares a same-mean shortcut
with the calendar-aware process, reports both a verdict and a calendar discount,
and maps the result to the next engineering test.}
\label{fig:audit-workflow}
\end{figure}

The audit record must be owned jointly.  Vulnerability management usually owns
MTTR, SLA reporting, remediation scope, ticket timing, and backlog evidence.
Change management owns release-window policy, change-freeze rules, approval
gates, and emergency-change procedures.  Platform, endpoint, cloud, or
application engineering owns deployment telemetry: release fractions, coverage
curves, rings, phases, rollback rules, and effective fielded coverage.  SOC and
incident response own detection, triage, escalation, validation, and emergency
response delays.  BAS, red-team, purple-team, and control-validation teams
provide evidence for the local residual-pressure ledger.  Risk, governance,
audit, or the CISO office owns the final interpretation, tolerance for
uncertainty, and policy action.  This ownership split matters because ticket
closure is not necessarily fielded coverage, an approved emergency change is not
necessarily deployed, and a written release policy is not evidence that
production assets actually updated on that schedule.

Table~\ref{tab:audit-record-compact} gives the minimum record.  The record
should be interval-first whenever possible.  Instead of recording only
\(L=2.70\), the analyst should record \(I_L=[L_-,L_+]\), the evidence used to
produce it, and the owner who signed off on that evidence.  The same principle
applies to \(m\), \(T\), \(\alpha\), cohort phases, \(\tau_\Sigma\), and the
local-channel rate scenario.  If input uncertainty is larger than the calendar
discount being tested, the correct output is an input-resolution finding, not a
precise capacity claim.

\begin{table}[H]
\centering
\caption{Minimum remediation-cadence audit record.  The audit requires ordinary
security-operations data to be recorded together before MTTR/SLA is used as
fielding evidence.}
\label{tab:audit-record-compact}
\small
\begin{tabularx}{\textwidth}{@{}p{0.23\textwidth}p{0.32\textwidth}X@{}}
\toprule
Field & Owner and evidence source & Purpose in the audit \\
\midrule

Scope and local channel &
Vulnerability management; asset owner; risk owner &
Defines the asset class, platform, control family, vulnerability class, or
technique family to which the claim applies. \\

Routine mean lag \(m\) &
VM tickets; scanner exports; endpoint or workload telemetry &
The value used by the continuous same-mean shortcut.  The event definition must
distinguish ticket closure from effective fielded coverage. \\

Release calendar \(T,\alpha\) &
Change calendar; release governance; deployment telemetry &
Records the release-window period \(T\) and the fraction \(\alpha\) of the
eligible routine backlog fielded at each window. \\

Rings and cohort phases &
Platform, endpoint, cloud, or application engineering &
Records whether the estate is updated synchronously, in rings, or in
phase-shifted cohorts. \\

Emergency-bypass rule &
Incident response; emergency change process; VM policy &
Identifies fixes that do not follow the routine cadence and should not be
averaged into the routine mean. \\

Non-fielding delay \(\tau_\Sigma\) &
SOC, validation, approval, packaging, and testing records &
Captures detection, triage, validation, approval, packaging, and other hard
delays not explained by the release train itself. \\

Residual-pressure interval \(I_L\) &
BAS, red-team, purple-team, control validation, incident postmortems, or expert
ledger &
Estimates local unresolved attacker-side pressure after existing controls,
detection, response, and compensating measures are accounted for. \\

Local-channel rate scenario &
Default normalized screening convention, local estimates, or scenario bands &
Declares the relative rates used to compute the capacity boundaries and calendar
discount.  If this scenario is uncertain, the audit reports a scenario
limitation. \\

Freshness and sign-off &
Risk, governance, audit, or CISO office &
Records observation window, last refresh date, uncertainty, evidence owner, and
governance interpretation. \\

\bottomrule
\end{tabularx}
\end{table}

Routine and emergency paths should be separated before interpreting any mean.  A
common dashboard failure is to report a single MTTR that mixes urgent exploited
vulnerabilities remediated through emergency change with routine issues waiting
for a release train.  Such an aggregate may be useful for high-level reporting,
but it does not answer the cadence question.  If two remediation paths have
materially different timing rules, approval rules, or coverage curves, they must
be recorded as separate channels before the audit verdict is used.  A combined
governance statement should be made only after the channel-specific findings are
known.

The residual-pressure score \(L\) is local.  It is not a generic CVSS, EPSS, KEV,
or severity score.  It summarizes the attacker-side pressure that remains for
the audited channel after the organization accounts for existing controls,
detection, response, hardening, and compensating measures.  In a mature audit,
\(I_L\) should be built from BAS campaigns, red-team or purple-team findings,
control-validation results, incident postmortems, and expert-scored evidence
mapped to the local control posture.  A small expert ledger is acceptable if it
states the technique family, defensive posture, score, uncertainty, evidence
source, and reviewer.  An unsupported point value chosen because it lies near a
threshold is not acceptable evidence.

Once the record is assembled, the calculation compares two representations of
the same routine remediation process.  The continuous same-mean shortcut treats
the routine mean lag as if defensive fielding catches up smoothly.  The
calendar-aware calculation uses the recorded release period, release fraction,
cohort geometry, emergency/routine split, hard delay, and local-channel rate
scenario.  The audit returns the mean-only boundary
\(L_{\mathrm{crit}}^{\mathrm{cont}}\), the calendar-aware boundary
\(L_{\mathrm{crit}}^{\mathrm{cal}}\), the discount
\(\Delta_{\mathrm{cal}}\), and the status in
Table~\ref{tab:audit-status-compact}.

\begin{table}[H]
\centering
\caption{Output statuses for a remediation-cadence audit.  The status should be
reported with input uncertainty, calendar discount, and evidence quality, not
only with point estimates.}
\label{tab:audit-status-compact}
\small
\begin{tabularx}{\textwidth}{@{}p{0.24\textwidth}p{0.38\textwidth}X@{}}
\toprule
Status & Governance meaning & Typical next action \\
\midrule

Mean-only adequate &
The residual-pressure interval is below both the same-mean and calendar-aware
boundaries, and the calendar discount is small relative to the claimed headroom.
&
Continue MTTR/SLA reporting, but retain cadence fields and rerun after process
changes. \\

Resolved cadence warning &
The residual-pressure interval is below the same-mean boundary but above the
calendar-aware boundary.  The release calendar changes the local capacity
verdict. &
Do not rely on MTTR/SLA alone.  Test shorter release windows, larger release
fractions, lower hard delay, or different ring phases. \\

Calendar-discount finding &
The residual-pressure evidence is too coarse for a sharp verdict, but the
calendar discount is material relative to the claimed residual-pressure headroom.
&
Disclose and test the release calendar before using the mean lag as fielding
evidence.  Improve \(I_L\) if a sharp verdict is required. \\

Outside under both &
The residual-pressure interval exceeds both boundaries.  Cadence may matter, but
it is not the main explanation for the failed capacity claim. &
Reduce residual pressure, hard delay, or control gaps. \\

Input-resolution limited &
Uncertainty in \(I_L\), \(m\), \(T\), \(\alpha\), phases, or \(\tau_\Sigma\) is
too large to resolve the cadence effect or the discount. &
Improve BAS, control-validation, deployment-coverage, or delay-budget evidence
before issuing a sharp cadence claim. \\

Phase, channel, or rate-scenario limitation &
The verdict or discount depends on ring phases, mixed routine/emergency paths,
or an undeclared local-channel rate scenario. &
Model rings explicitly, split emergency and routine channels, or run scenario
bands before making a governance claim. \\

\bottomrule
\end{tabularx}
\end{table}

The most important change in this workflow is the role of
\(\Delta_{\mathrm{cal}}\).  A narrow warning interval no longer makes the audit
useless.  It tells the team how much evidence resolution is required to defend a
mean-only fielding claim.  For example, a small calendar discount may be
evidence-resolution limited unless the BAS or control-validation ledger is
unusually sharp.  A larger discount, as in longer maintenance windows or coarse
release trains, can be operationally resolvable even with coarser residual-
pressure evidence.  Thus the audit can return a useful governance finding even
when the local residual-pressure interval is not precise enough for a binary
inside/outside verdict.

The immediate management output should be written narrowly:

\begin{quote}
For this local channel, this release calendar, this evidence interval, and this
declared rate scenario, MTTR/SLA reporting is adequate, calendar-sensitive,
input-resolution limited, or scenario-limited.  The next engineering test is to
change the release period, release fraction, hard-delay budget, channel split,
cohort phases, or evidence resolution.
\end{quote}

This compact workflow is the practical object used by the rest of the paper.
The next section applies it to worked audit packets.  Later sections define the
minimal calculation, explain why release windows change the capacity boundary,
and check the result under hard-delay placement and cohort-phasing
variations.

\section{Worked remediation-cadence audit packets}
\label{sec:worked-audit}

This section applies the audit workflow to notional but reproducible
remediation-cadence packets.  The purpose is not to claim empirical calibration
for a named enterprise.  Instead, the section separates what can be anchored by
public operational evidence from what must remain local.  Public sources can
support the timing side of the audit: monthly vendor release cycles, scheduled
maintenance windows, out-of-band release paths, exploited-vulnerability
remediation governance, and the general weakness of mean-only remediation
measurement \citep{nistPatchMgmt,msReleaseCycle,azureUpdateManager,cisaKEV,
cisaBOD2604,cyentiaP2P}.  Public sources cannot supply the residual-pressure
interval \(I_L\), because \(I_L\) depends on the audited organization's controls,
detection coverage, response procedures, compensating measures, and
BAS/red-team/control-validation evidence.

Table~\ref{tab:public-timing-calibration} summarizes this boundary.  It is
important for the method-paper framing of the article.  The worked packets below
use publicly motivated timing structures, but the residual-pressure ledger is
notional.  Production use must replace the notional ledger with a locally signed
evidence record.

\begin{table}[H]
\centering
\caption{Public evidence can anchor the timing side of a cadence audit, but not
the local residual-pressure interval \(I_L\).}
\label{tab:public-timing-calibration}
\small
\begin{tabularx}{\textwidth}{@{}p{0.25\textwidth}p{0.35\textwidth}X@{}}
\toprule
Public artifact & What it can anchor & What it cannot anchor \\
\midrule

Vendor release cycles &
Routine monthly security-update cadence and the existence of out-of-band release
paths. &
Enterprise deployment coverage, per-window release fraction, hard-delay budget,
or residual pressure. \\

Scheduled patching tools &
Recurring maintenance configurations, update schedules, machine scope, and
maintenance windows. &
Whether a particular estate achieved effective fielded coverage on the scheduled
cadence. \\

KEV and remediation directives &
Exploited-vulnerability governance, date-added and due-date objects, and
risk-based remediation context. &
Actual enterprise rollout, release-window geometry, or residual pressure after
local controls. \\

Remediation-measurement studies &
Long-tailed remediation behavior and the weakness of mean-only remediation
summaries. &
Calendar-aware capacity boundaries or local inside/outside capacity verdicts. \\

\bottomrule
\end{tabularx}
\end{table}

The local residual-pressure input is represented as an interval
\(I_L=[L_-,L_+]\).  For the worked packets, \(I_L\) is produced from a minimal
two-posture/two-technique ledger.  This ledger is not a general enterprise risk
matrix.  It is a local substitution ledger: rows are defender posture families,
columns are attacker technique families, and each cell records residual attacker
pressure after that defender posture is applied.  The example uses
\[
X_1=\hbox{patching-led posture},\qquad
X_2=\hbox{identity-led posture},
\]
\[
Y_1=\hbox{vulnerability exploitation},\qquad
Y_2=\hbox{valid-account abuse}.
\]
The scoring scale is \(0\) to \(2\).  Scores near \(0\) mean that BAS,
red-team, purple-team, control-validation, or postmortem evidence indicates that
the technique is reliably blocked, contained, or detected before material
progress.  Scores near \(1\) mean partial residual pressure.  Scores near \(2\)
mean that the technique remains repeatable under the audited posture or that
compensating controls are weak for the local channel.

The two-by-two convention converts the ledger into a scalar residual-pressure
input through the centered contrast
\begin{equation}
    s=\frac{S_{11}-S_{12}-S_{21}+S_{22}}{2},
    \qquad
    L=s^2 .
\label{eq:worked-centered-contrast}
\end{equation}
The centered contrast measures differential substitution pressure: whether the
attacker modes are differentially advantaged against the two defender postures.
The square removes the arbitrary sign of the contrast and gives the scalar
residual-pressure input used by the local-channel capacity calculation.  The
scale is not portable across enterprises.  It is meaningful only together with
the declared scoring rubric, local-channel convention, and threshold calculation
used in the audit.

The numerical cell values in Table~\ref{tab:worked-ledger} are selected to
instantiate a resolved cadence-sensitive packet for exposition; they are not
calibrated enterprise measurements.  Production use must replace these
demonstration cells with a locally signed BAS, red-team, control-validation,
postmortem, or expert-scored ledger and rerun the interval calculation.

\begin{table}[H]
\centering
\caption{Notional residual-pressure ledger for the worked packets.  Entries are
local residual-pressure scores after the defender posture is applied.  Intervals
show illustrative scoring uncertainty, not measurement precision from a real
enterprise dataset.}
\label{tab:worked-ledger}
\small
\begin{tabular}{@{}lcc@{}}
\toprule
 & \(Y_1\): vulnerability exploitation & \(Y_2\): valid-account abuse \\
\midrule

\(X_1\): patching-led posture &
\(0.20\;[0.195,0.205]\) &
\(1.84\;[1.83,1.85]\) \\

\(X_2\): identity-led posture &
\(1.85\;[1.84,1.86]\) &
\(0.20\;[0.195,0.205]\) \\

\bottomrule
\end{tabular}
\end{table}

Using the point values in Table~\ref{tab:worked-ledger},
\[
    s=\frac{0.20-1.84-1.85+0.20}{2}\approx -1.645,
    \qquad
    L\approx2.706 .
\]
Applying the same calculation to the illustrative score intervals gives
\[
    I_L=[2.657,2.756].
\]
The interval is reported immediately because the audit is not a third-decimal
forecast.  The question is whether the evidence interval is below both capacity
boundaries, above both, between them, or too wide relative to the calendar
discount.

The timing side of the worked packets uses a routine mean lag \(m=30\) days and
a non-fielding delay budget \(\tau_\Sigma=15\) days.  The hard-delay budget
captures detection, triage, validation, approval, testing, packaging, and other
latencies outside the routine release train.  Emergency bypass is excluded from
the routine mean.  If emergency fixes are part of the governance claim, they
require a separate fast-channel record before any combined statement is made.

Table~\ref{tab:worked-discount-screen} gives the calendar-discount screen for
the normalized local-channel rate scenario used in the running calculation.  The
continuous same-mean capacity boundary is \(2.767\).  The calendar-aware
boundary decreases as the same mean lag is implemented through coarser release
windows.  The calendar discount is the percentage of mean-only capacity consumed
by the release calendar.

\begin{table}[H]
\centering
\caption{Calendar-discount screen for the worked packets.  The routine mean lag
is held fixed at \(m=30\) days and the normalized hard delay is
\(\tau_\Sigma/m=0.5\).  Values are for the normalized local-channel rate
scenario used in the paper.}
\label{tab:worked-discount-screen}
\small
\begin{tabular}{@{}ccccc@{}}
\toprule
Cadence ratio \(T/m\) & Release period & Mean-matched \(\alpha\) &
\(L_{\mathrm{crit}}^{\mathrm{cal}}\) & Calendar discount \\
\midrule
0.5 & 15 days & 0.400 & 2.732 & 1.3\% \\
1.0 & 30 days & 0.667 & 2.623 & 5.2\% \\
1.5 & 45 days & 0.857 & 2.449 & 11.5\% \\
2.0 & 60 days & 1.000 & 2.286 & 17.4\% \\
\bottomrule
\end{tabular}
\end{table}

\begin{figure}[H]
\centering
\includegraphics[width=0.82\textwidth]{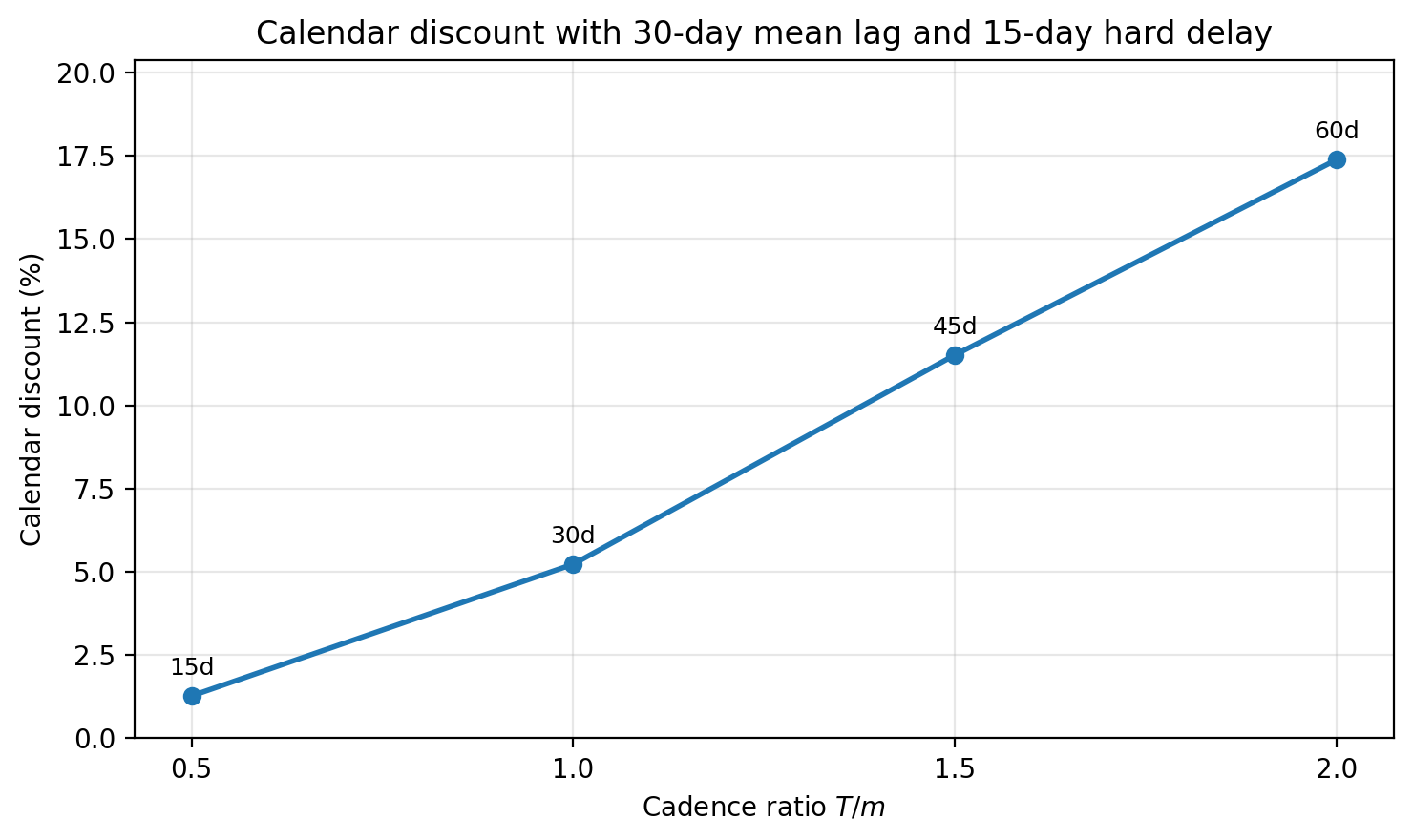}
\caption{Calendar discount under the normalized delayed scenario used in the
worked packets.  Coarser release windows consume a larger fraction of mean-only
local capacity, even when the reported mean lag is held fixed.}
\label{fig:worked-discount-screen}
\end{figure}

The headline worked packet uses the coarse two-month release train,
\(T/m=2\) and \(\alpha=1\).  This is a useful demonstration case because it
keeps the reported mean lag at 30 days while making the fielding calendar
material enough to be resolved with realistic evidence intervals.  The packet is
shown in Table~\ref{tab:worked-input-record}.  The same-mean shortcut classifies
the local channel as inside capacity because the entire interval
\(I_L=[2.657,2.756]\) lies below \(2.767\).  The bimonthly calendar-aware
calculation classifies the channel as outside capacity because the entire
interval lies above \(2.286\).  The calendar discount is \(17.4\%\).  The result
is therefore a resolved cadence warning: the mean remediation lag is not a safe
fielding representation for this local claim.

\begin{table}[H]
\centering
\caption{Headline worked packet: same reported mean lag, coarse routine release
calendar.  The packet is notional but reproducible; production use replaces the
ledger with local evidence.}
\label{tab:worked-input-record}
\small
\begin{tabularx}{\textwidth}{@{}p{0.29\textwidth}p{0.25\textwidth}X@{}}
\toprule
Field & Value & Audit interpretation \\
\midrule

Routine mean lag &
\(m=30\) days &
Measured from routine remediation eligibility to effective fielded coverage, not
only to ticket closure. \\

Routine release period &
\(T=60\) days &
Coarse two-month routine release train. \\

Release fraction &
\(\alpha=1\) &
All eligible routine backlog is fielded at each release window under the
same-mean convention. \\

Non-fielding delay budget &
\(\tau_\Sigma=15\) days &
Detection, triage, validation, approval, testing, packaging, and other hard
delays outside the release train. \\

Emergency channel &
Excluded from routine mean &
Emergency bypass has different timing rules and should be audited separately if
included in the governance claim. \\

Residual-pressure interval &
\(I_L=[2.657,2.756]\) &
Computed from the notional local substitution ledger in
Table~\ref{tab:worked-ledger}. \\

Continuous same-mean boundary &
\(L_{\mathrm{crit}}^{\mathrm{cont}}\approx2.767\) &
Mean-only capacity boundary under the normalized local-channel scenario. \\

Calendar-aware boundary &
\(L_{\mathrm{crit}}^{\mathrm{cal}}\approx2.286\) &
Capacity boundary under the bimonthly release calendar. \\

Calendar discount &
\(\Delta_{\mathrm{cal}}\approx17.4\%\) &
Fraction of mean-only capacity consumed by the recorded release calendar. \\

Audit status &
Resolved cadence warning &
The same packet is inside capacity under mean-only reporting but outside
capacity under the bimonthly release calendar. \\

\bottomrule
\end{tabularx}
\end{table}

The management reading is deliberately narrow:

\begin{quote}
For this local channel, evidence interval, hard-delay budget, and declared rate
scenario, the reported 30-day mean lag is not sufficient fielding evidence.  The
release calendar consumes a material share of mean-only capacity and changes the
local capacity verdict.
\end{quote}

The same residual-pressure ledger can also be used to explain why the monthly
case is different.  At \(T/m=1\), the calendar-aware boundary is
approximately \(2.623\) and the calendar discount is \(5.2\%\).  The stated
interval \(I_L=[2.657,2.756]\) still lies between the monthly and continuous
capacity boundaries, so the point packet remains a resolved warning.  However,
the smaller discount means that the verdict is more sensitive to scoring
resolution.  This is the practical value of reporting \(\Delta_{\mathrm{cal}}\):
a monthly cadence may be cadence-sensitive, but the organization needs a sharper
residual-pressure ledger to defend a binary inside/outside claim than it would
need for a coarser release train.

Table~\ref{tab:worked-ledger-sensitivity} makes this explicit.  Small
perturbations of the notional ledger leave the bimonthly calendar concern
visible, but quickly turn the monthly binary verdict into an input-resolution
question.  This is not a failure of the audit.  It is one of the intended audit
outputs: if the evidence is too coarse relative to the calendar discount, the
team should report an input-resolution or calendar-discount finding rather than
force a sharp verdict.

\begin{table}[H]
\centering
\caption{Sensitivity of the worked packets to residual-ledger uncertainty.  The
bimonthly case has a larger calendar discount; the monthly case is more
resolution-sensitive.}
\label{tab:worked-ledger-sensitivity}
\small
\begin{tabularx}{\textwidth}{@{}p{0.27\textwidth}p{0.22\textwidth}p{0.23\textwidth}X@{}}
\toprule
Residual-pressure treatment & Implied \(I_L\) &
Bimonthly \(T/m=2\) & Monthly \(T/m=1\) \\
\midrule

Stated score intervals &
\([2.657,2.756]\) &
Resolved cadence warning &
Resolved cadence warning, but with smaller evidence margin. \\

One off-diagonal score \(\pm0.02\) &
\([2.673,2.739]\) &
Resolved cadence warning &
Resolved cadence warning. \\

One off-diagonal score \(\pm0.05\) &
\([2.624,2.789]\) &
Calendar-discount finding; sharp verdict limited by upper tail &
Input-resolution limited near the continuous boundary. \\

Both off-diagonal scores \(\pm0.05\) &
\([2.544,2.873]\) &
Calendar-discount finding; sharp verdict limited by evidence width &
Input-resolution limited. \\

Broad local evidence interval &
\([2.50,2.90]\) &
Material calendar discount; improve \(I_L\) for a binary verdict &
Input-resolution limited. \\

\bottomrule
\end{tabularx}
\end{table}

The audit also supports a closed-loop engineering reading.  Starting from the
bimonthly packet, shortening the routine release train to a monthly cadence
reduces the calendar discount from \(17.4\%\) to \(5.2\%\), but does not fully
remove the warning for the stated evidence interval.  Shortening further to a
two-week cadence, \(T/m=0.5\), raises the calendar-aware boundary to
approximately \(2.732\) and reduces the discount to \(1.3\%\).  At that point
the point estimate \(L\approx2.706\) is inside capacity under the calendar-aware
calculation, but the interval \(I_L=[2.657,2.756]\) crosses the boundary.  The
right management conclusion is not that the intervention failed; it is that the
cadence lever has largely removed the timing discount, and the remaining
decision depends on residual-pressure evidence resolution or on reducing
residual pressure through control improvement.

\begin{table}[H]
\centering
\caption{Closed-loop reading of cadence interventions for the worked packet.
The action is not merely to issue a warning, but to identify which lever should
be tested next.}
\label{tab:worked-intervention}
\small
\begin{tabularx}{\textwidth}{@{}p{0.24\textwidth}p{0.20\textwidth}p{0.21\textwidth}X@{}}
\toprule
Routine release calendar & Calendar discount &
Status at \(I_L=[2.657,2.756]\) & Management reading \\
\midrule

Two-month train, \(T/m=2\) &
17.4\% &
Resolved cadence warning &
Mean-only reporting is unsafe for this local claim; cadence is a first-order
governance variable. \\

Monthly train, \(T/m=1\) &
5.2\% &
Resolved but resolution-sensitive warning &
Calendar still matters, but sharper local evidence is needed to defend a binary
verdict under moderate perturbations. \\

Two-week train, \(T/m=0.5\) &
1.3\% &
Threshold-adjacent / input-resolution limited &
Cadence discount is largely removed; remaining decision depends on improving
\(I_L\) or reducing residual pressure. \\

\bottomrule
\end{tabularx}
\end{table}

These worked packets establish three operational points before the model is
introduced.  First, two remediation processes can report the same 30-day mean
lag while consuming very different fractions of mean-only local capacity.
Second, the calendar discount is useful even when \(I_L\) is too coarse for a
sharp inside/outside verdict.  Third, the audit links a warning to an engineering
test: shorten the release period, increase the release fraction, separate
emergency and routine channels, reduce hard delay, compute cohort-phase effects,
or improve the local BAS/control-validation ledger.

The next section gives the minimal calculation that produces the capacity
boundaries used in these packets.  The following section explains why release
windows can change the verdict even when the mean lag is matched, and
Section~\ref{sec:validation-robustness} checks the release-geometry conclusions
under hard-delay placement, deployment rings, and cohort phasing.

\section{From the audit packet to calendar discount}
\label{sec:model}

The worked packets in Section~\ref{sec:worked-audit} used capacity boundaries
and calendar discounts before introducing the calculation that produces them.
This section defines that calculation as an audit interface.  It is not a SOC
simulator, a breach model, or a general theory of rollout kernels.  It specifies
what the backend consumes, what it returns, and how the returned quantities are
used in a remediation-cadence audit.  The local-channel state equations,
release-cycle construction, and characteristic-equation details are kept in
\ref{app:local-channel-backend}; the main text gives only the objects
needed to reproduce and interpret the audit.

The backend consumes the audit packet defined in Section~\ref{sec:audit}: the
routine mean lag \(m\), release-window period \(T\), per-window release fraction
\(\alpha\), cohort phases or ring geometry, emergency/routine channel split,
non-fielding delay budget \(\tau_\Sigma\), residual-pressure interval
\(I_L=[L_-,L_+]\), and the declared local-channel rate scenario.  The rate
scenario matters because the capacity boundaries are not universal constants.
The worked packets use a normalized screening convention in which local
defensive-revision and attacker-adjustment rates are measured in units of the
routine mean lag.  A production audit should either declare this screening
convention, supply local rate estimates, or run a scenario band over plausible
rates.  If the calendar discount changes materially across that band, the
correct audit output is a rate-scenario limitation rather than a sharp cadence
verdict.

The backend compares two fielding representations of the same routine
remediation process.  The first is the \emph{continuous same-mean shortcut}: the
routine mean lag is treated as if defensive fielding catches up smoothly.  This
is the representation implicitly used when a dashboard mean is treated as the
fielding process itself.  The second is the \emph{calendar-aware release
representation}: routine changes wait for release windows, and at each window
only the recorded fraction of the eligible backlog is fielded.  Rings and cohort
phases are handled as release geometry.  Emergency bypass is excluded from the
routine representation unless it is explicitly modeled as a separate channel.

For a synchronized routine release train, the same-mean comparison uses the
idealized fielding mean
\begin{equation}
    m_{\mathrm{cal}}
    =
    \frac{T}{2}
    +
    T\frac{1-\alpha}{\alpha}.
\label{eq:calendar-mean-lag}
\end{equation}
This formula has a simple operational reading.  The first term is the average
waiting time until the next release window under uniformly distributed arrival
phase.  The second term is the additional expected waiting created when each
window clears only a fraction \(\alpha\) of the remaining eligible backlog.  In
the worked packets, \(m_{\mathrm{cal}}\) is matched to the reported routine mean
lag so that the comparison isolates release geometry rather than average
remediation speed.  In a production audit, \(m\), \(T\), and \(\alpha\) should
come from deployment telemetry.  If the mean implied by coverage telemetry does
not match the reported MTTR, that mismatch is itself an audit finding.

The calculation returns four quantities, summarized in
Table~\ref{tab:backend-interface}.  The first is the mean-only capacity boundary,
\(L_{\mathrm{crit}}^{\mathrm{cont}}\).  The second is the calendar-aware capacity
boundary, \(L_{\mathrm{crit}}^{\mathrm{cal}}\), for the recorded release
geometry.  The third is the calendar discount, already introduced in
Section~\ref{sec:audit}.  The fourth is the position of the residual-pressure
interval \(I_L\) relative to the two boundaries.

\begin{table}[H]
\centering
\caption{Calculation interface used by the remediation-cadence audit.  The
backend is used to compute governance quantities, not to simulate the whole
enterprise.}
\label{tab:backend-interface}
\small
\begin{tabularx}{\textwidth}{@{}p{0.27\textwidth}p{0.30\textwidth}X@{}}
\toprule
Backend object & What it means & How the audit uses it \\
\midrule

\(L_{\mathrm{crit}}^{\mathrm{cont}}\) &
Mean-only capacity boundary under the continuous same-mean shortcut. &
Benchmark for the claim implicitly made when MTTR is treated as a fielding
model. \\

\(L_{\mathrm{crit}}^{\mathrm{cal}}\) &
Calendar-aware capacity boundary under the recorded release process. &
Tests whether release windows, release fraction, rings, hard delay, or channel
splitting change the local capacity verdict. \\

\(\Delta_{\mathrm{cal}}\) &
Fraction of mean-only local capacity consumed by calendarized fielding. &
Reports how much evidence resolution is needed before a mean-only fielding claim
is defensible. \\

Position of \(I_L\) &
Residual-pressure interval relative to both boundaries. &
Determines whether the finding is mean-only adequate, resolved cadence warning,
calendar-discount finding, outside under both, input-resolution limited, or
scenario-limited. \\

\bottomrule
\end{tabularx}
\end{table}

The most important audit case is no longer only a point estimate between two
thresholds.  A resolved cadence warning occurs when the residual-pressure
interval is below the mean-only boundary but above the calendar-aware boundary:
\[
    L_{\mathrm{crit}}^{\mathrm{cal}} < L_- \le L_+
    < L_{\mathrm{crit}}^{\mathrm{cont}} .
\]
In that case the same remediation packet is inside capacity under mean-only
reporting but outside capacity under the actual release calendar.  A
calendar-discount finding is weaker but still useful: \(I_L\) may be too wide
for a binary verdict, yet the discount may be material relative to the claimed
headroom.  This is the situation in which the audit tells the team that
MTTR/SLA should not be used alone unless release-calendar evidence is disclosed
and tested.

The calculation can also return an input-resolution or scenario limitation.  If
the uncertainty in \(I_L\), \(\alpha\), \(T\), cohort phases, or
\(\tau_\Sigma\) is wider than the calendar effect being tested, the audit should
not force a sharp inside/outside verdict.  If the result changes across
plausible local-channel rate scenarios, the audit should report a scenario band
rather than a single threshold.  These outcomes are not failures of the method.
They are governance findings: the organization lacks enough evidence resolution
to defend a mean-only fielding claim for the local channel.

The same interface also explains how to read engineering interventions.  Changing
the release period \(T\), the release fraction \(\alpha\), the hard-delay budget
\(\tau_\Sigma\), cohort phases, or the emergency/routine channel split changes
\(L_{\mathrm{crit}}^{\mathrm{cal}}\) and therefore
\(\Delta_{\mathrm{cal}}\).  Reducing residual attacker pressure through better
controls, detection, or response changes \(I_L\).  The audit is useful because it
keeps these two levers separate: one lever changes the fielding process, the
other changes the local residual-pressure evidence.

This interface is the only calculation layer needed in the main text.  The next
section explains why release windows can reduce the calendar-aware capacity
boundary even when the mean lag is matched.  \ref{app:local-channel-backend}
records the local-channel equations used by the backend, \ref{app:proof-calendarization}
gives the small-cadence result, \ref{app:same-mean-fielding-counterexample}
shows why same-mean fielding processes need not be equivalent, and
\ref{app:independent-dde-backend} summarizes the independent numerical
check used in Section~\ref{sec:validation-robustness}.

\section{Why release windows create a calendar discount}
\label{sec:calendarization-penalty}

Section~\ref{sec:model} defined the calculation interface: the audit packet is
converted into a mean-only capacity boundary, a calendar-aware capacity
boundary, a calendar discount, and an evidence-resolution status.  This section
explains why the calendar-aware boundary can be lower even when the reported
mean remediation lag is held fixed.  The detailed perturbation calculation is
given in \ref{app:proof-calendarization}; the main text keeps only the
operational mechanism and the quantities needed for the audit.

The mechanism is simple.  A continuous same-mean shortcut lets the fielded
defensive posture catch up at every instant.  A release calendar does not.  It
holds routine changes until the next window, then fields the recorded fraction
of the eligible backlog.  During the waiting part of the cycle, the estate
continues to operate with stale fielded posture.  During the release part of the
cycle, the correction arrives in synchronized batches.  Matching the average lag
does not match this timing response.  Two remediation processes can therefore
have the same reported mean lag while exposing the local channel to different
fielding dynamics.

For the normalized local-channel backend, synchronized release windows reduce
the calendar-aware capacity boundary initially by a second-order cadence effect:
\begin{equation}
    L_{\mathrm{crit}}^{\mathrm{cal}}(T)
    =
    L_{\mathrm{crit}}^{\mathrm{cont}}
    -
    \mathfrak c T^2
    +
    O(T^4),
    \qquad
    \mathfrak c>0 .
\label{eq:s5-small-cadence}
\end{equation}
This expression is not used to compute the finite worked-packet values.  It is a
local explanation of direction: when routine fielding is synchronized into
windows, the calendar-aware capacity boundary moves downward from the
continuous same-mean boundary.  The finite monthly, six-week, and bimonthly
values used in the audit are computed directly from the release-window backend.

The audit reports this loss as the calendar discount:
\begin{equation}
    \Delta_{\mathrm{cal}}
    =
    \frac{
    L_{\mathrm{crit}}^{\mathrm{cont}}
    -
    L_{\mathrm{crit}}^{\mathrm{cal}}
    }{
    L_{\mathrm{crit}}^{\mathrm{cont}}
    } .
\label{eq:s5-calendar-discount}
\end{equation}
The discount is not a breach probability and not a vulnerability score.  It is a
fielding-governance quantity: the fraction of mean-only local capacity consumed
by the recorded release calendar.  This is the quantity that lets the audit stay
useful when the residual-pressure interval is too coarse for a sharp binary
verdict.

To interpret the discount, the audit also records the width of the
residual-pressure interval relative to the mean-only boundary:
\begin{equation}
    W_L
    =
    \frac{L_+-L_-}{L_{\mathrm{crit}}^{\mathrm{cont}}}.
\label{eq:s5-evidence-width}
\end{equation}
When \(W_L\) is small relative to \(\Delta_{\mathrm{cal}}\), the evidence is
sharp enough to support a resolved inside/outside comparison.  When \(W_L\) is
comparable to or larger than \(\Delta_{\mathrm{cal}}\), the audit should usually
report an input-resolution or calendar-discount finding rather than force a
binary verdict.  This is not a failure of the method.  It is the method telling
the organization that its residual-pressure evidence is too coarse to justify
using MTTR as a fielding model.

Table~\ref{tab:s5-discount-reading} gives the practical reading used in this
paper.  The bands are not universal policy thresholds; they are an audit
language for relating timing loss to evidence resolution.  A small discount may
still matter for a high-assurance channel with sharp BAS/control-validation
evidence.  A larger discount should be treated as a first-order governance
variable even when residual-pressure evidence is coarse.

\begin{table}[H]
\centering
\caption{Practical reading of calendar discount.  The discount should be
interpreted together with the residual-pressure interval width \(W_L\) and the
declared local-channel rate scenario.}
\label{tab:s5-discount-reading}
\small
\begin{tabularx}{\textwidth}{@{}p{0.22\textwidth}p{0.38\textwidth}X@{}}
\toprule
Calendar discount & Evidence-resolution reading & Typical governance action \\
\midrule

Below about \(3\%\) &
Usually smaller than ordinary scoring and telemetry uncertainty, unless the
local ledger is unusually sharp. &
Report cadence as recorded; improve evidence if a sharp capacity verdict is
required. \\

About \(3\)--\(8\%\) &
Potentially material but resolution-sensitive.  A binary verdict requires a
well-supported residual-pressure interval. &
Disclose the release calendar and test whether BAS/control-validation evidence
can resolve the discount. \\

About \(8\)--\(15\%\) &
Often large enough to be visible with practical evidence intervals. &
Treat cadence as a governance variable; test shorter windows, larger release
fractions, or hard-delay reduction. \\

Above about \(15\%\) &
Calendarized fielding consumes a large share of mean-only capacity. &
Do not use MTTR/SLA alone as fielding evidence; run a calendar-aware audit before
making the local capacity claim. \\

\bottomrule
\end{tabularx}
\end{table}

For the normalized delayed scenario used in the worked packets,
\(\tau_\Sigma/m=0.5\) and
\(L_{\mathrm{crit}}^{\mathrm{cont}}\approx2.767\).  Holding the reported mean
lag fixed at 30 days, a two-week release screen has a discount of about \(1.3\%\),
a monthly release train about \(5.2\%\), a six-week train about \(11.5\%\), and a
bimonthly train about \(17.4\%\).  These values explain why the bimonthly packet
in Section~\ref{sec:worked-audit} is a cleaner headline demonstration than the
monthly packet.  The monthly case is operationally common and still important,
but its discount is narrow enough that a sharp binary verdict requires unusually
good residual-pressure evidence.  The bimonthly case has the same reported mean
lag but a much larger discount, so the calendar effect remains visible with
coarser evidence intervals.

The same logic explains the closed-loop intervention reading.  If a bimonthly
routine release train is shortened to monthly, the discount falls substantially
but may remain resolution-sensitive.  If it is shortened further, the calendar
discount may become small relative to the residual-pressure uncertainty; at that
point the cadence lever has largely removed the timing loss, and the remaining
decision depends on improving the local ledger or reducing residual pressure
through control changes.  The audit therefore separates two engineering levers:
change the fielding process to move \(L_{\mathrm{crit}}^{\mathrm{cal}}\), or
change the security posture to move \(I_L\).

Hard delay does not remove this interpretation.  Detection, validation,
approval, packaging, and testing delays reduce the available local capacity in
both the same-mean and calendar-aware representations.  The release calendar
then determines how much additional capacity is consumed by batching routine
fielding into windows.  Section~\ref{sec:validation-robustness} checks that the
calendar gap is not an artifact of assigning the same hard-delay budget to one
side of the feedback loop.

The discount also depends on the declared local-channel rate scenario.  The
worked packets use the normalized screening convention, but production audits
should either declare that convention or run a scenario band over plausible
local rates.  If the discount remains in the same practical band across the
scenario range, the cadence finding is robust.  If the discount moves from, for
example, a negligible band to a material band when the rate scenario changes,
the correct output is a rate-scenario limitation.  This is why the rate scenario
is part of the audit packet rather than a hidden modeling choice.

Finally, calendar discount is not eliminated by saying that a release train uses
rings.  Rings and cohort phases change the calendar-aware release geometry, and
therefore can change \(L_{\mathrm{crit}}^{\mathrm{cal}}\) and
\(\Delta_{\mathrm{cal}}\).  They do not automatically recover the continuous
same-mean shortcut.  The next section validates this point numerically: adding
equal-phase cohorts at fixed per-asset cadence approaches a phase-averaged
calendar process, not the continuous benchmark, and cohort staggering can help
or hurt near capacity.

The practical result of this section is the audit rule used throughout the rest
of the paper:

\begin{quote}
Do not ask only whether a point estimate of \(L\) lies between two thresholds.
Report the calendar discount and compare it with the residual-pressure evidence
resolution.  If the discount is material relative to the claimed headroom or the
evidence width, MTTR/SLA should not be used alone as a fielding representation.
\end{quote}

\ref{app:proof-calendarization} gives the small-cadence derivation.
\ref{app:same-mean-fielding-counterexample} shows independently why
same-mean fielding processes need not be equivalent.  Section~\ref{sec:validation-robustness}
then checks the finite running values, hard-delay placement, and cohort-geometry
effects used by the audit.

\section{Validation and release-geometry checks}
\label{sec:validation-robustness}

The worked packets and the calendar-discount interpretation rely on finite
release-window boundaries, not on the small-cadence approximation alone.  This
section checks that the practical findings are not artifacts of one numerical
implementation, one hard-delay placement, or one rate scenario.  The checks are
reported as security-operations validation rather than as a numerical-methods
section.  The technical backend uses an exact one-cycle release map for the
single-total-delay case and collocation cross-checks for alternative delay
placements; implementation details are summarized in
Appendix~\ref{app:independent-dde-backend}.

The validation addresses five questions.  First, do the finite calendar-aware
boundaries used in the audit stabilize under the independent backend?  Second,
is the calendar-discount reading robust over a plausible attacker-adjustment rate
band?  Third, does the calendar discount survive alternative placements of the
same hard-delay budget?  Fourth, does splitting a release train into more
cohorts recover the continuous same-mean benchmark?  Fifth, is cohort staggering
always beneficial?  The answers are: yes, qualitatively yes for the practical
bands used here, yes, no, and no.  These are the release-geometry facts a
security team needs before relying on MTTR/SLA as fielding evidence.

The monthly point is the most resolution-sensitive worked case because its
calendar discount is only about \(5.2\%\) under the normalized screening
scenario.  It is therefore a useful numerical stress test: if the backend is
unstable, this narrow case would be the first to move.  For the normalized
delayed scenario,
\[
    T/m=1,\qquad
    \alpha=2/3,\qquad
    \tau_\Sigma/m=0.5,
\]
the continuous same-mean boundary is
\[
    L_{\mathrm{crit}}^{\mathrm{cont}}=2.767423\ldots .
\]
Table~\ref{tab:s6-monthly-threshold-check} reports the independent
calendar-aware boundary as the backend resolution is increased.  The boundary
stabilizes at approximately \(2.623\).  Thus the monthly discount used in the
worked packet,
\[
    \Delta_{\mathrm{cal}}
    =
    \frac{2.767423-2.622676}{2.767423}
    \approx 5.2\% ,
\]
is not a visible artifact of the backend resolution.

\begin{table}[H]
\centering
\caption{Independent numerical check for the monthly calendar-aware boundary
\((T/m=1,\alpha=2/3,\tau_\Sigma/m=0.5)\).  The continuous same-mean boundary is
\(2.767423\ldots\).}
\label{tab:s6-monthly-threshold-check}
\small
\begin{tabular}{@{}cc@{}}
\toprule
Backend resolution & \(L_{\mathrm{crit}}^{\mathrm{cal}}\) \\
\midrule
4  & 2.622481 \\
6  & 2.622681 \\
8  & 2.622678 \\
10 & 2.622677 \\
12 & 2.622676 \\
\bottomrule
\end{tabular}
\end{table}

\begin{figure}[H]
\centering
\includegraphics[width=0.82\textwidth]{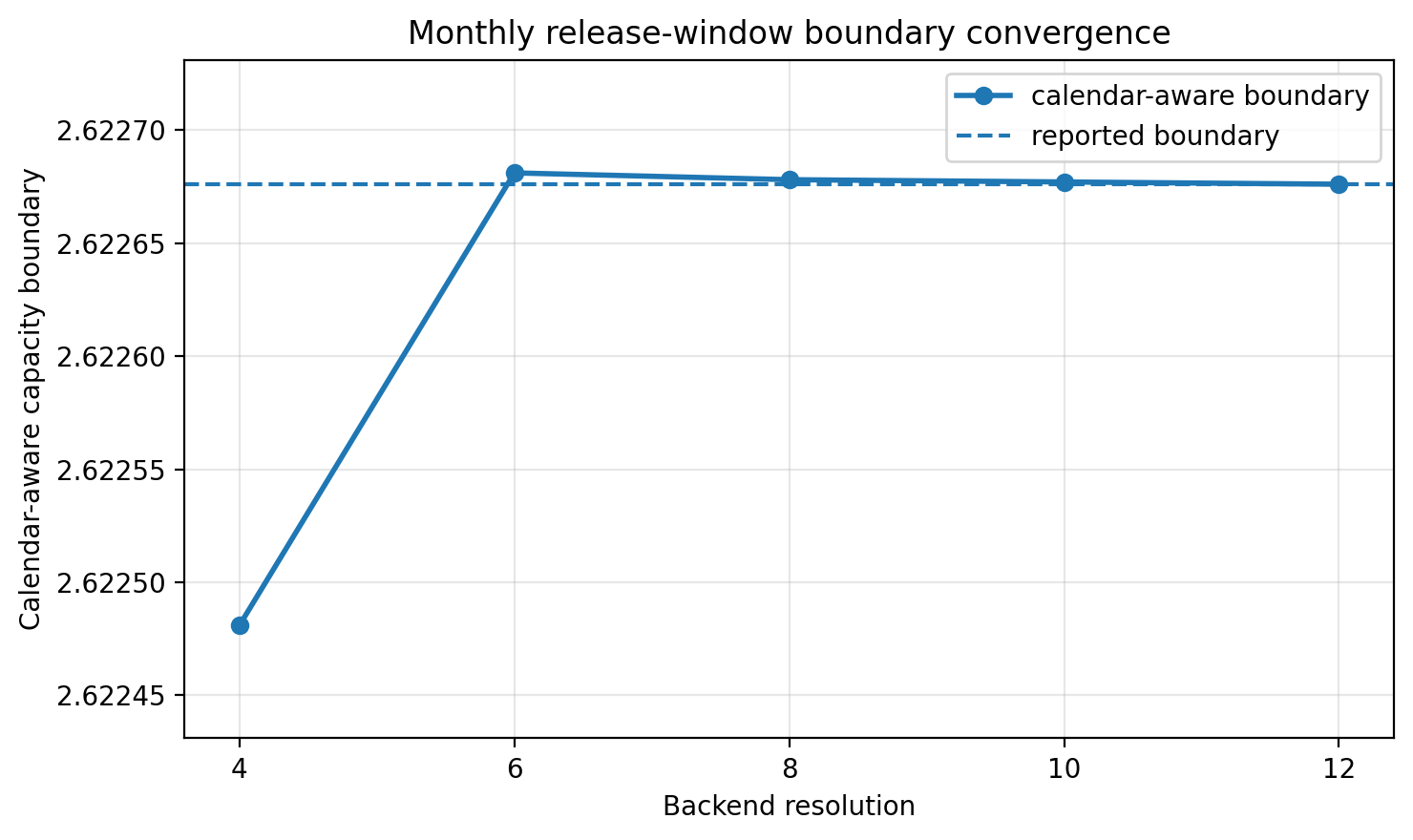}
\caption{Independent boundary check for the monthly audit point.  The
calendar-aware boundary stabilizes near \(2.623\), below the continuous
same-mean boundary \(2.767\).}
\label{fig:s6-monthly-threshold-check}
\end{figure}

The second check concerns the rate scenario.  The worked packets use the
normalized screening convention \(\kappa=1\), but the audit record requires the
rate scenario to be declared rather than hidden.  Table~\ref{tab:s6-rate-band}
therefore reports the calendar-discount band over a 16-fold range of
attacker-adjustment rates, \(\kappa\in[0.25,4]\), with the same normalized
hard-delay budget \(\tau_\Sigma/m=0.5\).  The qualitative reading is stable.  The
two-week screen remains below \(2\%\), the monthly train stays in the
resolution-sensitive \(3\)--\(8\%\) band, and the two-month train remains
material throughout the band, never falling below about \(12\%\).  The
bimonthly discount is not monotone in \(\kappa\), so production audits should
report a scenario band rather than a single point when local rates are uncertain.

\begin{table}[H]
\centering
\caption{Rate-scenario band for calendar discount.  Values use
\(\tau_\Sigma/m=0.5\) and the same-mean release screen.  The band supports the
paper's qualitative reading: short cadence is negligible, monthly cadence is
resolution-sensitive, and two-month cadence is material across the tested range.}
\label{tab:s6-rate-band}
\small
\begin{tabular}{@{}cccccc@{}}
\toprule
\(\kappa\) & \(T/m=0.5\) & \(T/m=1.0\) & \(T/m=1.5\) & \(T/m=2.0\) & \(L_{\mathrm{crit}}^{\mathrm{cont}}\) \\
\midrule
0.25 & 0.86\% & 3.44\% & 7.79\% & 13.39\% & 4.2795 \\
0.50 & 1.04\% & 4.20\% & 9.44\% & 15.65\% & 3.1497 \\
1.00 & 1.27\% & 5.23\% & 11.52\% & 17.38\% & 2.7674 \\
2.00 & 1.53\% & 6.52\% & 13.33\% & 16.35\% & 2.7972 \\
4.00 & 1.75\% & 7.88\% & 13.61\% & 12.42\% & 2.9994 \\
\bottomrule
\end{tabular}
\end{table}

The third check concerns the non-fielding delay budget.  In the worked packets,
\(\tau_\Sigma/m=0.5\) represents detection, triage, validation, approval,
testing, packaging, and other delays outside the routine release train.  A
reviewer may ask whether the calendar gap is created by placing that hard delay
on one side of the local channel.  The independent verification answers this
more strongly than a split table: for the monthly packet, placing the same total
delay on the fielded-posture-to-attacker link, sampling the intended posture at
\(t_n-\tau_\Sigma\), or placing the delay on either feedback direction gives the
same calendar-aware boundary to less than \(10^{-5}\).  In all cases the monthly
boundary is \(2.622676\ldots\).  The calendar discount is therefore a release-
calendar effect, not a bookkeeping artifact of where the same total hard-delay
budget is placed.

The fourth check concerns deployment rings.  A common operational response to a
cadence warning is to say that the release train is not synchronized because it
uses rings.  Rings can be useful, but they do not automatically justify the
continuous same-mean shortcut.  At fixed per-asset cadence, increasing the
number of equal-phase cohorts converges to a phase-averaged calendar process,
not to continuous fielding.  Each asset still sees a release calendar.

Table~\ref{tab:s6-dense-staggering-monthly} shows the effect for the monthly
case.  Equal-phase cohorts move the calendar-aware boundary slightly upward,
from about \(2.623\) to about \(2.628\).  That is a real improvement, but it
does not recover the continuous same-mean boundary \(2.767423\ldots\).  The
calendar discount remains about \(5\%\).  The operational conclusion is direct:
a team cannot defend mean-only reporting merely by saying that monthly release
is split into rings.  The ring geometry must be recorded and checked.

\begin{table}[H]
\centering
\caption{Equal-phase staging at fixed monthly per-asset cadence does not recover
the continuous same-mean benchmark.  Values use \(T/m=1\),
\(\alpha=2/3\), and \(\tau_\Sigma/m=0.5\).}
\label{tab:s6-dense-staggering-monthly}
\small
\begin{tabular}{@{}ccc@{}}
\toprule
Release geometry & Calendar-aware boundary & Calendar discount \\
\midrule
Synchronized monthly release, \(R=1\) & 2.623 & 5.2\% \\
Equal-phase cohorts, \(R=2\) & 2.628 & 5.0\% \\
Equal-phase cohorts, \(R=4\) & 2.628 & 5.0\% \\
Equal-phase cohorts, \(R=8\) & 2.628 & 5.0\% \\
Dense equal-phase calendar limit & 2.628 & 5.0\% \\
Continuous same-mean shortcut & 2.767 & 0.0\% \\
\bottomrule
\end{tabular}
\end{table}

\begin{figure}[H]
\centering
\includegraphics[width=0.82\textwidth]{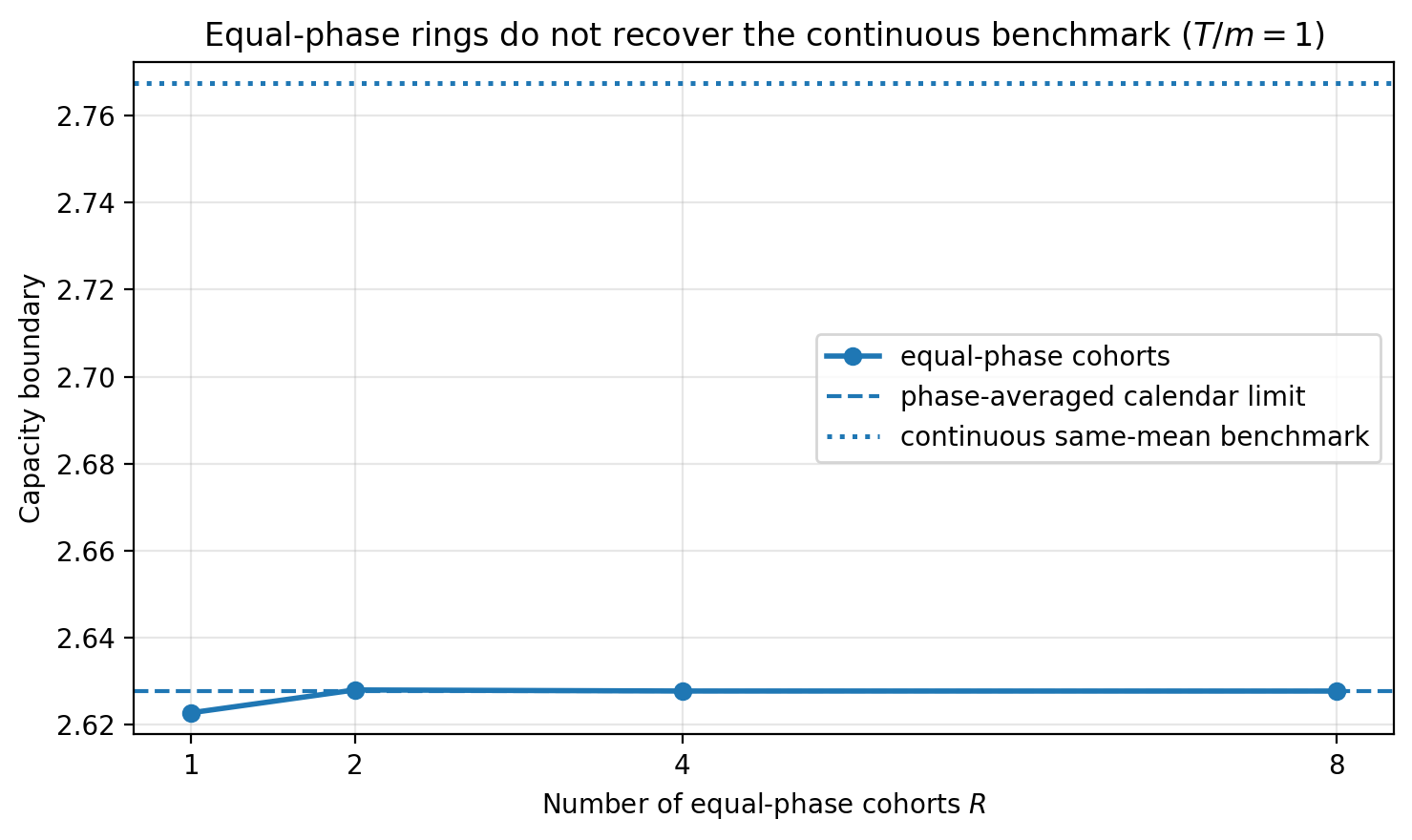}
\caption{Increasing the number of equal-phase cohorts at fixed monthly
per-asset cadence converges to a phase-averaged calendar boundary, not to the
continuous same-mean benchmark.}
\label{fig:s6-dense-staggering-monthly}
\end{figure}

The fifth check is whether staggering is always stabilizing.  It is not.  Cohort
phasing is a release-geometry variable, not a universal remedy.  Near a capacity
boundary, different phase schedules can move the local channel in different
directions.  Table~\ref{tab:s6-phase-geometry} gives two near-boundary examples.
In the first, equal staggering turns a small positive growth rate into a small
negative one.  In the second, equal staggering turns a small negative growth rate
into a positive one.  The point is not that staging is bad.  The point is that
the phase schedule must be treated as evidence, not as a verbal substitute for
the continuous shortcut.

\begin{table}[H]
\centering
\caption{Cohort phase geometry can help or hurt near capacity.  Growth is the
local calendar-aware release-cycle growth rate; negative means inside capacity
and positive means outside capacity.}
\label{tab:s6-phase-geometry}
\small
\begin{tabular}{@{}lcccc@{}}
\toprule
Case & \(T/m\) & \(L\) & Synchronized growth &
Equal-staggered growth, \(R=4\) \\
\midrule
Staggering helps & 1.5 & 2.456 & \phantom{-}0.000703 & -0.000490 \\
Staggering hurts & 1.8 & 2.342 & -0.000138 & \phantom{-}0.000593 \\
\bottomrule
\end{tabular}
\end{table}

\begin{figure}[H]
\centering
\includegraphics[width=0.86\textwidth]{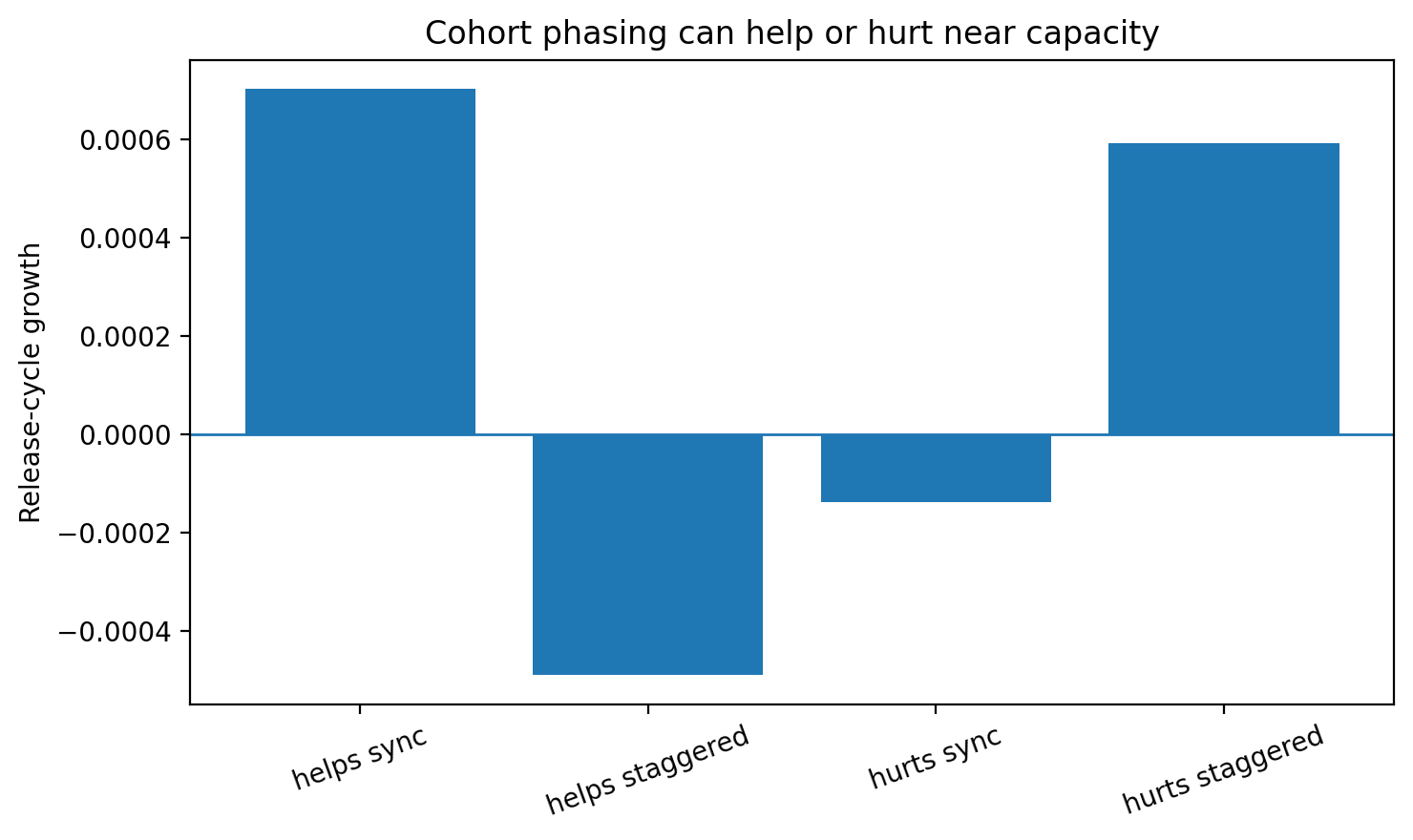}
\caption{Cohort phasing is a regime-dependent design variable.  Staggering can
improve or worsen the local verdict when the channel is close to capacity.}
\label{fig:s6-phase-geometry}
\end{figure}

Together, the checks support the audit interpretation without overclaiming.  The
finite monthly boundary is numerically stable; the discount bands are stable in
practical terms across a broad \(\kappa\) scenario range; the calendar boundary
is invariant to tested placements of the same total hard-delay budget; deployment
rings at fixed per-asset cadence do not recover the continuous same-mean
benchmark; and cohort staggering can help or hurt near capacity.  These are
fielding-governance findings.  They tell a security team when release-calendar
evidence must be recorded and tested before MTTR/SLA is used as a fielding
model.

\section{Related work and positioning}
\label{sec:related-work}

This paper sits at the intersection of enterprise vulnerability management,
security measurement, remediation governance, adversary emulation, and timing
models.  Its contribution is not a new vulnerability-severity score, a new
exploit-prediction model, or a general adaptive-defense theory.  The contribution
is narrower: a remediation-cadence audit that tests whether a mean remediation
metric is a safe fielding representation, and that reports the calendar discount
and evidence-resolution status when release geometry is hidden by MTTR/SLA
reporting.

Enterprise patch-management guidance provides the operational setting.  NIST
frames enterprise patch management as a preventive-maintenance process involving
identification, prioritization, acquisition, installation, and verification of
patches \citep{nistPatchMgmt}.  Public vendor and cloud documentation shows that
routine security updates, scheduled maintenance windows, recurring patching
configurations, and out-of-band update paths are ordinary operational objects
rather than theoretical constructs \citep{msReleaseCycle,azureUpdateManager}.
These sources motivate the timing fields in the audit packet: release-window
period, release fraction, maintenance window, emergency bypass, and effective
fielded coverage.

Security-metrics work motivates the caution against mean-only reporting.
Quantified security metrics are useful for governance, but prior surveys and
critical reviews emphasize that aggregate measures can be weakly validated,
context-dependent, and difficult to interpret operationally
\citep{verendel2009,pendletonMetrics}.  Vulnerability-management studies and
industry remediation analyses also show that remediation data can be skewed,
long-tailed, censored, or sensitive to the definition of closure
\citep{cyentiaP2P}.  The present audit follows this measurement tradition but
focuses on one hidden variable: release cadence.  The issue is not only whether
the mean remediation lag is high or low; it is whether the mean is a defensible
fielding model for the local security claim being made.

Vulnerability scoring, exploitability signals, and known-exploited-vulnerability
programs are adjacent but not substitutes for the residual-pressure input used
here.  CVSS provides a standardized severity framework \citep{firstCVSS}; EPSS
and related work estimate exploitation likelihood or help prioritize remediation
\citep{jacobsEPSS,jacobsBetterExploit}; empirical studies compare vulnerability
severity with observed exploitation \citep{allodiMassacci,holmAfridiCVSS}; and
remediation-selection work addresses how to choose which vulnerabilities to fix
under constraints \citep{shahRemediation}.  CISA's KEV catalog and risk-based
remediation directives illustrate governance based on known exploitation and
remediation deadlines \citep{cisaKEV,cisaBOD2604}.  These signals are important
for scope and prioritization, but they do not describe how fixes are fielded
through release windows, rings, change freezes, or emergency bypass paths.  They
also do not provide a local residual-pressure interval after an organization's
specific controls, detections, response processes, and compensating measures are
taken into account.

The residual-pressure ledger is instead connected to adversary-emulation and
control-validation practice.  MITRE ATT\&CK provides a shared vocabulary for
techniques and defensive observations \citep{mitreATTACK}.  CALDERA and Atomic
Red Team illustrate repeatable adversary-emulation and ATT\&CK-mapped testing
practices \citep{apacheCaldera,atomicRedTeam}.  NIST SP 800-53A provides a
general assessment frame for evaluating whether controls are implemented and
effective \citep{nist80053a}.  The audit does not depend on any one tool or
framework.  It requires the local ledger to identify the technique family,
defender posture, score interval, evidence source, observation window, and
reviewer.  This is why \(L\) is treated as a local evidence interval rather than
as a public CVE-derived score.

The work also relates to moving-target defense, deception, timing games, and
adaptive security models.  FlipIt-style games and moving-target defense surveys
study the timing of attack and defense, reconfiguration, and strategic
adaptation \citep{flipit,choMTDsurvey,senguptaMTDsurvey,pawlickDeception}.  Work
on patching dynamics and epidemic-style remediation models studies how updates
or security actions propagate through populations \citep{eshghiPatching,
taynitskiyImpulseSIR}.  These literatures establish that timing and adaptation
matter.  The present paper addresses a more specific enterprise-operations
question: when a team reports remediation through a mean lag, does the release
calendar consume enough local capacity that the mean should not be used as the
fielding model?

Finally, sampled-data, networked-control, and delay-system methods provide the
technical backend used to compute the capacity boundaries and release-geometry
checks \citep{zhangNCS,hetelSwitched,guDelay,fridmanDelay,breda2005,bredaBook,
breda2022}.  This backend explains why same-mean fielding processes need not be
equivalent and why synchronized release windows can create a calendar discount.
However, the article's security contribution is not the backend itself.  The
contribution is the audit interface: an operational packet, a same-mean versus
calendar-aware comparison, a calendar discount, an evidence-resolution reading,
and a governance status that tells enterprise teams when release-calendar
evidence must be disclosed and tested before MTTR/SLA is used as fielding
evidence.

\section{Limitations and responsible use}
\label{sec:limitations}

The audit is intentionally limited.  It is a local fielding diagnostic, not a
breach prediction, exploit-probability model, CVE prioritizer, or complete
enterprise risk model.  A calendar-discount finding means that the release
calendar consumes a material part of mean-only local capacity, or that the
available evidence is too coarse to justify treating MTTR as a fielding model.
It does not mean that an incident will occur, that a specific vulnerability
should be patched first, or that calendarized release trains are generally
unsafe.

The worked packets are not empirical calibrations on a named enterprise dataset.
The timing side is motivated by public operational artifacts such as release
cycles, scheduled patching, maintenance windows, out-of-band update paths, KEV
governance, and remediation-measurement studies.  The residual-pressure ledger,
however, is notional.  This is appropriate for a method paper, but it limits the
claim: the paper demonstrates how the audit is run, how a calendar discount is
reported, and how evidence resolution changes the interpretation.  It does not
estimate how frequently such findings occur across enterprises.

The residual-pressure interval \(I_L\) is the most important local input and the
largest source of governance risk.  It depends on score construction, evidence
quality, defender-posture taxonomy, technique grouping, and the mapping from BAS,
red-team, purple-team, control-validation, postmortem, or expert-scored evidence
to the audited local channel.  The two-by-two ledger used in the worked packets
is a minimal substitution ledger, not a universal enterprise risk matrix.
Production use should report the scoring rubric, evidence sources, observation
window, reviewer, sign-off authority, and an interval rather than an unsupported
point value.

The calendar discount also depends on the declared local-channel rate scenario.
The worked packets use a normalized screening convention.  This convention is
useful for demonstrating the audit and for comparing release geometries, but it
is not a calibrated enterprise forecast.  A production audit should either
declare the normalized convention, supply local rate estimates, or run scenario
bands over plausible defensive-revision and attacker-adjustment rates.  If the
discount or status changes materially across the scenario band, the appropriate
output is a rate-scenario limitation rather than a single sharp verdict.

The remediation timing inputs can also be imperfect.  MTTR may be censored,
right-tailed, asset-dependent, or based on ticket closure rather than effective
fielded coverage.  Release fraction \(\alpha\) may differ across platforms,
asset classes, and maintenance windows.  Cohort phases and ring policies may
differ between policy documents and actual deployment telemetry.  Emergency and
routine remediation paths may be mixed in one dashboard mean even though they
follow different timing, approval, testing, and coverage rules.  The audit is
only as reliable as the deployment-coverage and delay-budget evidence used to
populate the packet.

The model itself is local and deliberately simplified.  The main calculation
treats one local adaptation channel and a declared routine cadence.  It does not
model strategic attacker anticipation of release calendars, adversarial
manipulation of emergency-bypass rules, correlated enterprise-wide failures,
multi-channel coupling, asset-dependent exploit chains, or mode-dependent
cadence in which different attacker techniques and defender postures have
different fielding laws.  These are important extensions, but including them in
the main audit would make the procedure less usable for security operations.

Responsible use therefore requires conservative interpretation.  A resolved
cadence warning should trigger engineering tests of release period, release
fraction, hard-delay budget, cohort phases, emergency/routine channel separation,
and residual-pressure reduction.  A calendar-discount finding should trigger
disclosure and testing of the release calendar before MTTR/SLA is used as
fielding evidence.  An input-resolution finding should trigger better BAS,
control-validation, deployment-coverage, or delay-budget evidence.  These
statuses are not failures of the audit.  They are the audit's intended governance
outputs.

\section{Conclusion}
\label{sec:conclusion}

MTTR and SLA compliance are useful remediation metrics, but they are not
fielding models.  They summarize how long remediation takes on average; they do
not describe how fixes actually reach the estate through release windows,
maintenance periods, rings, partial backlog clearing, hard delays, and emergency
bypass paths.  This paper introduced a remediation-cadence audit for that gap.
The audit records the routine mean lag, release period, release fraction, cohort
geometry, emergency/routine channel split, non-fielding delay budget,
residual-pressure interval, and local-channel rate scenario, then compares a
continuous same-mean shortcut with the recorded release calendar.

The main output is the calendar discount: the fraction of mean-only local
capacity consumed by calendarized routine fielding.  This shifts the audit away
from a fragile point comparison and toward an evidence-resolution question.  If
the discount is small relative to the residual-pressure evidence width, the
right finding may be input-resolution limited.  If the discount is material
relative to the claimed headroom, MTTR/SLA should not be used alone as fielding
evidence.  If the residual-pressure interval is resolved across the two capacity
boundaries, the audit returns a cadence warning.

The worked packets show how this reading changes the operational message.  With
the same reported 30-day mean lag and the same non-fielding delay budget, a
coarse two-month release train consumes about 17.4\% of mean-only local capacity,
while a monthly train consumes about 5\% and a shorter two-week screen about
1\%.  The coarser case is easier to resolve with practical residual-pressure
evidence; the monthly case is more resolution-sensitive.  This distinction is
the point of the audit.  It tells a security team not only whether a point
estimate falls between two boundaries, but also how much evidence resolution is
needed before a mean-only fielding claim is defensible.

The release-geometry checks add two practical findings.  Splitting a monthly
release train into equal-phase deployment rings at fixed per-asset cadence does
not recover the continuous same-mean benchmark; it approaches a phase-averaged
calendar process.  Cohort staggering is also not automatically stabilizing near
capacity.  It can help or hurt depending on the local channel and phase
schedule.  Rings and cohort phases therefore belong in the audit packet rather
than in a verbal assurance that rollout is ``staged.''

The practical implication is simple.  A vulnerability-management program should
not treat MTTR/SLA as complete fielding evidence when the release calendar can
consume a material share of local capacity.  The team should record cadence,
release fraction, rings, emergency bypass, hard delay, residual-pressure
evidence, and rate scenario; compute the calendar discount; compare it with
evidence resolution; and choose the next engineering test.  The audit does not
replace vulnerability prioritization or incident response.  It gives enterprise
security teams a reproducible way to decide when the remediation calendar itself
must be disclosed and tested before a mean remediation metric is used in
security-governance claims.

\section*{Data and code availability}

The worked packets use a notional residual-pressure ledger and do not contain
enterprise-confidential data.  A supplementary reproducibility package is
provided with the manuscript.  It includes an audit-packet template, a
residual-pressure ledger template, the notional worked packet, the notional
ledger, source data for the figures and tables, a verdict-classification script,
and an exact running-case backend script that recomputes the continuous boundary,
finite calendar-aware boundaries, calendar discounts, and the \(\kappa\) scenario
band reported in the paper.

The supplementary scripts reproduce the audit reading and the reported
running-case checks; they do not infer residual pressure from public CVE, CVSS,
EPSS, KEV, OSV, or advisory data.  Public sources are used only to anchor timing
and governance context.  Production use must replace the notional ledger with
local BAS, red-team, control-validation, postmortem, or expert-scored evidence
and should report uncertainty intervals and rate-scenario assumptions.

\appendix

\section{Local-channel backend equations}
\label{app:local-channel-backend}

This appendix records the local-channel backend used by the audit calculation.
The main text treats the calculation as an interface from an audit packet to
capacity boundaries and calendar discount.  This appendix gives the state
representation used to compute those quantities.

All timing quantities can be nondimensionalized by the routine mean lag \(m\).
The worked packets use the normalized screening convention
\[
    m=1,\qquad \mu_X=\mu_Y=\kappa=1,
\]
with the non-fielding delay reported as \(\tau_\Sigma/m\).  A production audit
may instead supply local rate estimates or run a scenario band over plausible
rate ratios.  The rate scenario is therefore part of the audit packet rather
than a hidden modeling choice.

Let \(a(t)\) denote the intended defensive revision and \(x(t)\) the defensive
posture actually fielded on the estate.  Let \(b(t)\) denote the attacker's
intended technique adjustment and \(y(t)\) the technique share actually in use.
Before inserting the separate non-fielding delay budget, the local-channel
backend is
\begin{equation}
\begin{aligned}
\dot a(t)&=-\mu_X a(t)+\sqrt L\,y(t),\\
\dot b(t)&=-\mu_Y b(t)-\sqrt L\,x(t),\\
\dot y(t)&=\kappa\bigl(b(t)-y(t)\bigr).
\end{aligned}
\label{eq:app-local-channel-flow}
\end{equation}
The fielding rule for \(x(t)\) is the component changed by the audit comparison.

The continuous same-mean shortcut treats fielding as a smooth catch-up process,
\begin{equation}
    \dot x(t)=\frac{1}{m}\bigl(a(t)-x(t)\bigr).
\label{eq:app-continuous-fielding}
\end{equation}
For this representation, without non-fielding delay, the local capacity boundary
is obtained from
\begin{equation}
    D(z)+\kappa L=0,
    \qquad
    D(z)=(z+\mu_X)(z+\mu_Y)(z+\kappa)(1+mz).
\label{eq:app-cont-characteristic}
\end{equation}
With a total non-fielding delay budget \(\tau_\Sigma\), the corresponding
single-total-delay characteristic equation is
\begin{equation}
    e^{z\tau_\Sigma}
    (z+\mu_X)(z+\mu_Y)(z+\kappa)(1+mz)
    +\kappa L=0.
\label{eq:app-cont-characteristic-delay}
\end{equation}
The continuous same-mean boundary \(L_{\mathrm{crit}}^{\mathrm{cont}}\) is the
value of \(L\) at which the principal root has zero real part.

The synchronized release-window representation holds routine fielding between
release windows and updates the fielded posture at release times \(t_n=nT\):
\begin{equation}
    \dot x(t)=0,\qquad t\ne t_n,
\label{eq:app-between-releases}
\end{equation}
\begin{equation}
    x(t_n^+)=(1-\alpha)x(t_n^-)+\alpha a(t_n^-).
\label{eq:app-calendar-release-rule}
\end{equation}
Rings and cohort phases apply the same release rule to cohorts at their recorded
phases.  Emergency bypass is not included in this routine rule unless the
emergency path is explicitly modeled as a separate channel.

For the no-delay synchronized case, write \(s=(a,b,x,y)^\top\).  Between release
windows,
\begin{equation}
\dot s(t)=A_Ls(t),
\qquad
A_L=
\begin{pmatrix}
-\mu_X&0&0&\sqrt L\\
0&-\mu_Y&-\sqrt L&0\\
0&0&0&0\\
0&\kappa&0&-\kappa
\end{pmatrix},
\label{eq:app-AL}
\end{equation}
and the release update is
\begin{equation}
J_\alpha=
\begin{pmatrix}
1&0&0&0\\
0&1&0&0\\
\alpha&0&1-\alpha&0\\
0&0&0&1
\end{pmatrix}.
\label{eq:app-Jalpha}
\end{equation}
The one-window release-cycle matrix is
\begin{equation}
    M_L(T)=J_\alpha e^{A_LT}.
\label{eq:app-release-cycle-matrix}
\end{equation}
The no-delay calendar-aware capacity boundary is obtained from
\begin{equation}
    \rho\!\left(M_L(T)\right)=1,
\label{eq:app-rho-boundary}
\end{equation}
where \(\rho(\cdot)\) is spectral radius.  Equivalently, the release-cycle
growth rate is
\begin{equation}
    g_{\mathrm{cal}}(T,L)=\frac{1}{T}\log \rho\!\left(M_L(T)\right),
\label{eq:app-growth-rate-no-delay}
\end{equation}
and the boundary satisfies \(g_{\mathrm{cal}}(T,L_{\mathrm{crit}}^{\mathrm{cal}})=0\).

With positive non-fielding delay, the same idea is implemented by the independent
delay-equation backend summarized in \ref{app:independent-dde-backend}.
The finite monthly, six-week, and bimonthly thresholds reported in the main text
are computed from the release-window backend, not from the small-cadence
asymptotic formula.

For an idealized synchronized release process with uniformly distributed arrival
phase and geometric clearing of the remaining eligible backlog, the mean
fielding lag is
\begin{equation}
    m_{\mathrm{cal}}=\frac{T}{2}+T\frac{1-\alpha}{\alpha}.
\label{eq:app-calendar-mean}
\end{equation}
This is the same-mean convention used in the worked packets.  Production audits
should estimate \(m\), \(T\), and \(\alpha\) from deployment telemetry and treat
any mismatch between reported MTTR and empirical coverage timing as an audit
finding.

\section{Small-cadence release-window penalty}
\label{app:proof-calendarization}

This appendix gives the perturbation result behind the calendar-discount
interpretation.  The main text uses only the operational consequence:
synchronized release windows can lower the calendar-aware capacity boundary even
when the mean fielding lag is matched.  The finite thresholds used in the worked
packets are computed directly; the result below explains the direction of the
effect for small release periods.

Let \(N\) be the fielding-update direction
\begin{equation}
N=
\begin{pmatrix}
0&0&0&0\\
0&0&0&0\\
1&0&-1&0\\
0&0&0&0
\end{pmatrix},
\qquad
N^2=-N.
\label{eq:app-N}
\end{equation}
The release update can be written as
\[
    J_\alpha=I+\alpha N=\exp(\beta N),
    \qquad
    \beta=-\log(1-\alpha).
\]
The continuous same-mean generator is
\[
    G_0=A_L+\frac{1}{m}N,
\]
which corresponds to the smooth fielding law
\(\dot x=(a-x)/m\).

Assume that the continuous same-mean backend has a simple principal critical
point
\[
    z_0=i\omega_0,\qquad \omega_0>0,
\]
at branch strength \(L_0\), so that
\begin{equation}
    D(i\omega_0)+\kappa L_0=0,
    \qquad
    D'(i\omega_0)\ne0,
\label{eq:app-principal-crossing}
\end{equation}
where
\[
    D(z)=(z+\mu_X)(z+\mu_Y)(z+\kappa)(1+mz).
\]
Define
\begin{equation}
    \Psi=\frac{D'(z_0)}{D(z_0)}.
\label{eq:app-Psi}
\end{equation}
For the synchronized release rule with mean matching
\begin{equation}
    m=\frac{T}{2}+T\frac{1-\alpha}{\alpha},
    \qquad
    \alpha(T)=\frac{T}{m+T/2},
\label{eq:app-mean-matching}
\end{equation}
the local calendar-aware capacity boundary has the expansion
\begin{equation}
    L_{\mathrm{crit}}^{\mathrm{cal}}(T)
    =
    L_0-\mathfrak cT^2+O(T^4),
    \qquad T\to0,
\label{eq:app-main-expansion}
\end{equation}
where
\begin{equation}
    \mathfrak c
    =
    L_0\left(
    \operatorname{Re}q+
    \operatorname{Im}q\,
    \frac{\operatorname{Im}\Psi}{\operatorname{Re}\Psi}
    \right),
    \qquad
    q=\frac{\omega_0^2}{12(1+im\omega_0)}.
\label{eq:app-c-coefficient}
\end{equation}
For the principal simple crossing of the positive first-order local-channel
family, \(\mathfrak c>0\).

To derive this expansion, first note that the mean-matching rule gives
\[
    \beta(T)
    =
    -\log(1-\alpha(T))
    =
    \log\frac{m+T/2}{m-T/2}
    =
    \frac{T}{m}+\frac{T^3}{12m^3}+O(T^5).
\]
The one-window product \(J_{\alpha(T)}e^{A_LT}\) is similar to the symmetric
section
\begin{equation}
    S_T=e^{\beta(T)N/2}e^{A_LT}e^{\beta(T)N/2}.
\label{eq:app-symmetric-section}
\end{equation}
Because this symmetric representation is reversible in \(T\), the release-cycle
generator
\[
    G_T=\frac{1}{T}\log S_T
\]
has an even expansion around the continuous generator:
\begin{equation}
    G_T=G_0+T^2G_2+O(T^4).
\label{eq:app-generator-expansion}
\end{equation}
The corresponding determinant expansion is
\begin{equation}
    m\det(zI-G_T)
    =
    D(z)+\kappa L
    +T^2\widetilde C_2(z,L)+O(T^4).
\label{eq:app-det-expansion}
\end{equation}
At the continuous critical point \(z_0=i\omega_0\), where
\(D(z_0)+\kappa L_0=0\), the second-order term reduces to
\begin{equation}
    \widetilde C_2(z_0,L_0)
    =
    -\frac{\kappa L_0 z_0^2}{12(1+mz_0)}
    =
    \kappa L_0
    \frac{\omega_0^2}{12(1+im\omega_0)}
    =
    \kappa L_0 q.
\label{eq:app-C2-hopf}
\end{equation}

Continue the critical point as
\begin{equation}
    z_T=i(\omega_0+\nu T^2)+O(T^4),
    \qquad
    L_T=L_0+\ell T^2+O(T^4),
\label{eq:app-perturbation}
\end{equation}
with \(\nu,\ell\in\mathbb R\).  Substituting into
\eqref{eq:app-det-expansion} and taking order-\(T^2\) terms gives
\begin{equation}
    D'(z_0)i\nu+\kappa\ell+\kappa L_0q=0.
\label{eq:app-linearized}
\end{equation}
Since \(D'(z_0)=\Psi D(z_0)=-\kappa L_0\Psi\), this is equivalent to
\begin{equation}
    -L_0\Psi i\nu+\ell+L_0q=0.
\label{eq:app-linearized-Psi}
\end{equation}
Taking imaginary and real parts yields
\begin{equation}
    \nu=
    \frac{\operatorname{Im}q}{\operatorname{Re}\Psi},
    \qquad
    \ell=
    -L_0\left(
    \operatorname{Re}q+
    \operatorname{Im}q\,
    \frac{\operatorname{Im}\Psi}{\operatorname{Re}\Psi}
    \right).
\label{eq:app-ell-nu}
\end{equation}
Thus \(L_T=L_0-\mathfrak cT^2+O(T^4)\).

For the normalized no-delay local channel
\[
    \mu_X=\mu_Y=\kappa=m=1,
\]
the continuous characteristic equation is
\[
    (1+z)^4+L=0.
\]
The principal critical point is \(z_0=i\), the continuous boundary is \(L_0=4\),
and the synchronized same-mean release calendar satisfies
\begin{equation}
    L_{\mathrm{crit}}^{\mathrm{cal}}(T)
    =
    4-\frac{1}{3}T^2+O(T^4).
\label{eq:app-normalized-result}
\end{equation}

With positive non-fielding delay \(\tau_\Sigma\), the continuous characteristic
factor becomes
\begin{equation}
    D_\tau(z)
    =
    e^{z\tau_\Sigma}
    (z+\mu_X)(z+\mu_Y)(z+\kappa)(1+mz).
\label{eq:app-Dtau}
\end{equation}
The same local perturbation applies with
\[
    \Psi
    \quad\text{replaced by}\quad
    \Psi_\tau
    =
    \frac{D_\tau'(z_0)}{D_\tau(z_0)}
    =
    \tau_\Sigma
    +\frac{1}{z_0+\mu_X}
    +\frac{1}{z_0+\mu_Y}
    +\frac{1}{z_0+\kappa}
    +\frac{m}{1+mz_0}.
\]
The finite delayed thresholds in the main text are computed directly from the
release-window backend and checked independently in
\ref{app:independent-dde-backend}.

\section{Same-mean fielding processes can differ}
\label{app:same-mean-fielding-counterexample}

This appendix records a simple reason why MTTR is not a fielding model.  Two
fielding processes can have the same mean lag while presenting different timing
responses to a local adaptation channel.

Let a linear fielding process be represented by a nonnegative lag distribution
with transform
\begin{equation}
    H(z)=\mathbb E[e^{-zD}],
\label{eq:app-lag-transform}
\end{equation}
where \(D\) is the fielding delay.  The mean lag is
\[
    \mathbb E[D]=m.
\]
Matching \(m\) matches only the first derivative of \(H(z)\) at the origin:
\[
    H(z)=1-mz+O(z^2).
\]
It does not determine the higher-order timing response.

For example, a pure deterministic delay \(D=m\) has
\begin{equation}
    H_{\mathrm{delay}}(z)=e^{-mz}
    =
    1-mz+\frac{m^2z^2}{2}+O(z^3).
\label{eq:app-pure-delay-transform}
\end{equation}
The continuous first-order catch-up shortcut with the same mean has
\begin{equation}
    H_{\mathrm{cont}}(z)=\frac{1}{1+mz}
    =
    1-mz+m^2z^2+O(z^3).
\label{eq:app-first-order-transform}
\end{equation}
Both have the same mean lag \(m\), but they have different second-order timing
responses.  At oscillatory frequencies \(z=i\omega\), they also have different
phase and amplitude responses.  A local feedback channel can therefore see two
same-mean fielding processes differently.

The synchronized release-window model used in the audit gives another example.
Under the idealized same-mean convention, the fielding delay can be written as
\[
    D=U+KT,
\]
where \(U\) is the waiting time to the next release window and is uniform on
\([0,T]\), while \(K\) is the number of additional windows required before the
eligible item is fielded.  If each window clears a fraction \(\alpha\) of the
remaining eligible backlog, then
\[
    \mathbb P(K=k)=\alpha(1-\alpha)^k,\qquad k=0,1,2,\ldots .
\]
The transform is
\begin{equation}
    H_{\mathrm{cal}}(z)
    =
    \frac{1-e^{-zT}}{zT}
    \cdot
    \frac{\alpha}{1-(1-\alpha)e^{-zT}}.
\label{eq:app-calendar-transform}
\end{equation}
Its mean is
\begin{equation}
    -H_{\mathrm{cal}}'(0)
    =
    \frac{T}{2}+T\frac{1-\alpha}{\alpha}.
\label{eq:app-calendar-transform-mean}
\end{equation}
Thus \(H_{\mathrm{cal}}\) can be mean-matched to the continuous shortcut, but it
is not the same transfer function.  Calendarized fielding preserves release
phase, batching, and partial backlog clearing that are invisible in MTTR.

This counterexample is the reason the audit compares fielding representations
rather than comparing mean lags alone.  A reported MTTR can be correct as an
average and still be incomplete as evidence for how defensive changes reach the
estate.

\section{Independent numerical backend}
\label{app:independent-dde-backend}

This appendix summarizes the independent numerical backend used to check the
finite delayed boundaries and release-geometry results reported in
Section~\ref{sec:validation-robustness}.  The purpose is reproducibility of the
reported audit quantities, not enterprise calibration of the residual-pressure
ledger.

For the single-placement total-delay check, the backend is exact rather than a
collocation approximation.  In the normalized delayed local channel, the
fielded posture \(x\) is piecewise constant between release windows.  If the
total non-fielding delay \(\tau_\Sigma\) is placed on the fielded-posture input
to attacker adjustment, then \(x(t-\tau_\Sigma)\) has one breakpoint per release
cycle when \(0<\tau_\Sigma\le T\).  The one-cycle evolution can therefore be
written as a closed finite-dimensional map built from matrix exponentials.

For \(m=1\), \(\mu_X=\mu_Y=1\), and attacker-adjustment rate \(\kappa\), let the
continuous state during a cycle be \((a,b,y)\) and augment it with the current
and previous fielded-posture values \((x_{\mathrm{cur}},x_{\mathrm{prev}})\).
During the first interval of length \(\tau_\Sigma\), the delayed input is
\(x_{\mathrm{prev}}\); during the remaining interval of length
\(T-\tau_\Sigma\), the delayed input is \(x_{\mathrm{cur}}\).  With
\(r=\sqrt L\), the affine subsystem for \((a,b,y)\) is generated by
\begin{equation}
B_L=
\begin{pmatrix}
-1&0&r\\
0&-1&0\\
0&\kappa&-\kappa
\end{pmatrix},
\qquad
f_L=
\begin{pmatrix}0\\-r\\0\end{pmatrix},
\label{eq:app-exact-Bf}
\end{equation}
where \(f_L\) multiplies the delayed fielded-posture input.  The two affine
matrix exponentials over \(\tau_\Sigma\) and \(T-\tau_\Sigma\) give an exact
one-cycle map on \((a,b,y,x_{\mathrm{cur}},x_{\mathrm{prev}})\).  At the release
window, the fielding update is
\begin{equation}
    x_{\mathrm{cur}}^+=(1-\alpha)x_{\mathrm{cur}}^-+\alpha a(T^-),
    \qquad
    x_{\mathrm{prev}}^+=x_{\mathrm{cur}}^- .
\label{eq:app-exact-release}
\end{equation}
Let \(\mathcal M_L\) denote the resulting exact one-cycle matrix.  The
calendar-aware growth rate is
\begin{equation}
    g_{\mathrm{cal}}(T,L)=\frac{1}{T}\log\rho(\mathcal M_L),
\label{eq:app-dde-growth}
\end{equation}
and the calendar-aware capacity boundary solves
\begin{equation}
    g_{\mathrm{cal}}(T,L_{\mathrm{crit}}^{\mathrm{cal}})=0.
\label{eq:app-dde-boundary}
\end{equation}

The continuous same-mean boundary used for comparison is computed from
\begin{equation}
    e^{z\tau_\Sigma}(1+z)^3(\kappa+z)+\kappa L=0,
\label{eq:app-cont-kappa-delay}
\end{equation}
where the principal imaginary-axis crossing gives
\(L_{\mathrm{crit}}^{\mathrm{cont}}\).  For the normalized screening scenario
\(\kappa=1\), \(\tau_\Sigma/m=0.5\), this gives
\(L_{\mathrm{crit}}^{\mathrm{cont}}=2.767423\ldots\).  The exact release-cycle
map gives the finite calendar-aware boundaries reported in the main text:
\begin{center}
\begin{tabular}{@{}cccc@{}}
\toprule
\(T/m\) & \(\alpha\) & \(L_{\mathrm{crit}}^{\mathrm{cal}}\) & Calendar discount \\
\midrule
0.5 & 0.400 & 2.732235 & 1.27\% \\
1.0 & 0.667 & 2.622676 & 5.23\% \\
1.5 & 0.857 & 2.448731 & 11.52\% \\
2.0 & 1.000 & 2.286408 & 17.38\% \\
\bottomrule
\end{tabular}
\end{center}

Alternative hard-delay placements were checked separately.  Placing the same
total delay on the fielded-posture-to-attacker link, sampling the intended
posture at \(t_n-\tau_\Sigma\), and collocation checks with delay placed on the
other feedback directions all give the same monthly boundary to less than
\(10^{-5}\):
\[
    L_{\mathrm{crit}}^{\mathrm{cal}}(T/m=1,\alpha=2/3,\tau_\Sigma/m=0.5)
    =2.622676\ldots .
\]
This supports the interpretation in Section~\ref{sec:validation-robustness}:
the calendar discount is a release-calendar effect rather than a bookkeeping
artifact of where the same total hard delay is placed.

For cohort-phasing checks, the release update is applied to cohorts at their
recorded phases rather than to the whole estate at one synchronized instant.
At fixed per-asset cadence, increasing the number of equal-phase cohorts
converges to a phase-averaged calendar operator.  It does not converge to the
continuous same-mean shortcut, because each asset still experiences calendarized
fielding.  This is the numerical backend for the ring and staggering findings in
Section~\ref{sec:validation-robustness}.

The supplementary reproducibility package supplied with the manuscript contains
the notional audit packet, residual-pressure ledger, source data for the figures
and tables, a verdict-classification script, and an exact running-case backend
script.  The backend script recomputes the continuous boundary, finite
calendar-aware boundaries, calendar discounts, and \(\kappa\) scenario band.  It
does not infer residual-pressure scores from public CVE, CVSS, EPSS, KEV, OSV,
or advisory data.  Production use must replace the notional ledger with local
BAS, red-team, control-validation, postmortem, or expert-scored evidence and
must declare the rate scenario used for threshold and discount calculations.

\end{document}